\documentclass[12pt]{article}
\pdfoutput=1

\usepackage[bulletsep]{collref}
\usepackage{amssymb,graphicx}
\usepackage[intlimits]{amsmath}
\usepackage{bbm}
\usepackage[small]{subfigure}
\usepackage{xcolor}
\usepackage{MnSymbol}
\usepackage{arydshln}

\makeatletter \@addtoreset{equation}{section} \makeatother

\makeatletter
\let\old@startsection=\@startsection
\let\oldl@section=\l@section
\renewcommand{\@startsection}[6]{\old@startsection{#1}{#2}{#3}{#4}{#5}{#6\mathversion{bold}}}
\renewcommand{\l@section}[2]{\oldl@section{\mathversion{bold}#1}{#2}}
\makeatother

\makeatletter
\let\old@makecaption=\@makecaption
\def\@makecaption{\small\old@makecaption}
\makeatother

\renewcommand{\geq}{\geqslant}

\newcommand{\Bst}{{\rm{Bst}}}

\begin{document}

\thispagestyle{empty}
\begin{flushright}\footnotesize
\vspace{0.6cm}
\end{flushright}

\renewcommand{\thefootnote}{\fnsymbol{footnote}}
\setcounter{footnote}{0}

\begin{center}
{\Large\textbf{\mathversion{bold} 't Hooft loops and integrability}

\par}

\vspace{0.8cm}

\textrm{Charlotte~Kristjansen$^{1}$ and
Konstantin~Zarembo$^{1,2}$}
\vspace{4mm}

\textit{${}^1$Niels Bohr Institute, Copenhagen University, Blegdamsvej 17, \\ 2100 Copenhagen, Denmark}\\
\textit{${}^2$Nordita, KTH Royal Institute of Technology and Stockholm University,
Hannes Alfv\'{e}ns V\"{a}g 12, 114 19 Stockholm, Sweden}\\
\vspace{0.2cm}
\texttt{kristjan@nbi.dk, zarembo@nordita.org}

\vspace{3mm}


\par\vspace{1cm}

\textbf{Abstract} \vspace{3mm}

\begin{minipage}{13cm} 

We consider the defect CFT defined by a 't Hooft line embedded in ${\cal N}=4$ super Yang-Mills theory.  By explicitly quantizing around
the given background we exactly reproduce a prediction from S-duality for the correlators between the 't Hooft line and 
chiral primaries in the bulk and pave the way for higher loop analyses for non-protected operators.  
Furthermore, we demonstrate at the leading perturbative order that 
correlators  between the 't Hooft line and non-protected bulk operators can be efficiently computed using integrability. As a byproduct we find new integrable overlaps in $\mathfrak{sl}(2)$ spin chains in different representations.

\end{minipage}
\end{center}

\vspace{0.5cm}

\newpage
\setcounter{page}{1}
\renewcommand{\thefootnote}{\arabic{footnote}}
\setcounter{footnote}{0}

\newcommand{\CC}{\mathbbm{C}}
\newcommand{\kkb}{{\color{teal} \mathring{a}}}
\newcommand{\knb}{{\color{blue} \o}}
\newcommand{\nkb}{{\color{blue} \o}}
\newcommand{\nnb}{{\color{violet} \ae}}
\newcommand{\Yvec}{{\bf Y}_{J\ell M}^{(q)}(\theta,\phi)}


\section{Introduction}

A 't~Hooft loop \cite{tHooft:1977nqb} is a typical disorder operator defined by singular boundary conditions for gauge fields, rather than by taking a product of field operators. In essence the 't~Hooft loop is a trajectory of a point-like Dirac monopole. When embedded in  ${\cal N}=4$ super Yang Mills theory, the 't~Hooft line constitutes a 1/2 BPS defect of co-dimension three and gives rise to  a defect conformal field theory~\cite{Kapustin:2005py}.
This defect CFT has been subject of extensive analyses building on S-duality and supersymmetric localization \cite{Gomis:2009ir,Gomis:2009xg,Gomis:2011pf},  but should be amenable to other exact tools such as the boundary conformal bootstrap and integrability. 

The holographic dual of the 't Hooft line consists  of a D1-brane appropriately embedded in the $AdS_5\times S^5$ background~\cite{Diaconescu:1996rk}, and the boundary conditions provided by this object for the string sigma model belong to the list of integrable ones given  in~\cite{Dekel:2011ja}. Furthermore, the boundary bootstrap equations for the present set-up were argued to immediately follow from  those of the more studied co-dimension one, 1/2 BPS defect version of ${\cal N}=4$ SYM, dual to the  D3-D5 probe brane system~\cite{Liendo:2016ymz}. Seemingly, the 't Hooft line set-up is the simplest defect CFT where all these exact methods come together. 

 The  't Hooft loop is related to the more studied Wilson loop by S-duality~\cite{Montonen:1977sn,Witten:1978ma,Osborn:1979tq}. A number of exact results obtained for the
 Wilson loop using matrix model techniques can hence be directly translated to the 't Hooft loop. This holds for instance for  the expectation value of the 't Hooft loop itself as well as its correlators with chiral primary operators in the bulk. We shall demonstrate that exact results for correlators with non-protected local bulk operators can be obtained from integrability. Our approach
 is for the moment perturbative and should be seen as a necessary preparatory step for a full non-perturbative solution evoking integrability  bootstrap in the spirit of~\cite{Buhl-Mortensen:2017ind,Komatsu:2020sup,Gombor:2020kgu,Gombor:2020auk}. 
 
 At the leading perturbative order the computation of correlators reduces to a combinatorial problem which can be handled efficiently using the tools of integrability~\cite{deLeeuw:2015hxa,Buhl-Mortensen:2015gfd}. At the quantum level, the computations entail the study of beautiful examples of exactly solvable quantum mechanical problems, including Dirac's original mo\-no\-pole problem~\cite{Dirac:1931kp} and the problem of  a spin one particle coupled to a scalar in a monopole potential. A first outcome of the perturbative analysis is a solid confirmation of  the predictions made by S-duality and a number of exact expressions for the leading contribution to correlators between
 the 't Hooft line and non-protected operators in specific subsectors.
 
 Our paper is organized as follows. We begin by introducing the  't Hooft line in section~\ref{kinematics} and turn to discussing the implications of S-duality for its expectation value and its correlators with local, protected  bulk operators in section~\ref{S-duality}. Section~\ref{QFT} is devoted to the quantization of the 't Hooft line and ends with the above mentioned 
confirmation of an S-duality prediction.  A discussion of how to reproduce the S-duality prediction from supergravity
and string theory has been relegated to an appendix. 
In section~\ref{integrability} we demonstrate the leading order 
integrability of the 't Hooft line. Finally,
section~\ref{conclusion} contains our conclusion.

\section{Kinematics \label{kinematics}}

The 't~Hooft loop represents a trajectory of a point-like monopole, and is defined by the  prescribed singularity of the gauge fields. For the static 't~Hooft line
\cite{Kapustin:2005py}:
\begin{eqnarray}
 F_{ij }&=&\frac{B}{2}\,\varepsilon _{ijk}\,\frac{x_k }{r^3}, \label{F_cl}
 \\
 \Phi _I&=&\frac{Bn_I}{2r}\,.
 \label{Phi_cl}
\end{eqnarray}
 These should be regarded as boundary conditions  for the fields in the path integral. For the straight 't~Hooft line, $x_i$ are the spatial coordinates, more generally they should be regarded as coordinates of the 3d hyperplane orthogonal to the loop at a given point. And the boundary conditions are imposed at $x_i\rightarrow 0$.
 
 \begin{figure}[t]
 \centerline{\includegraphics[width=3cm]{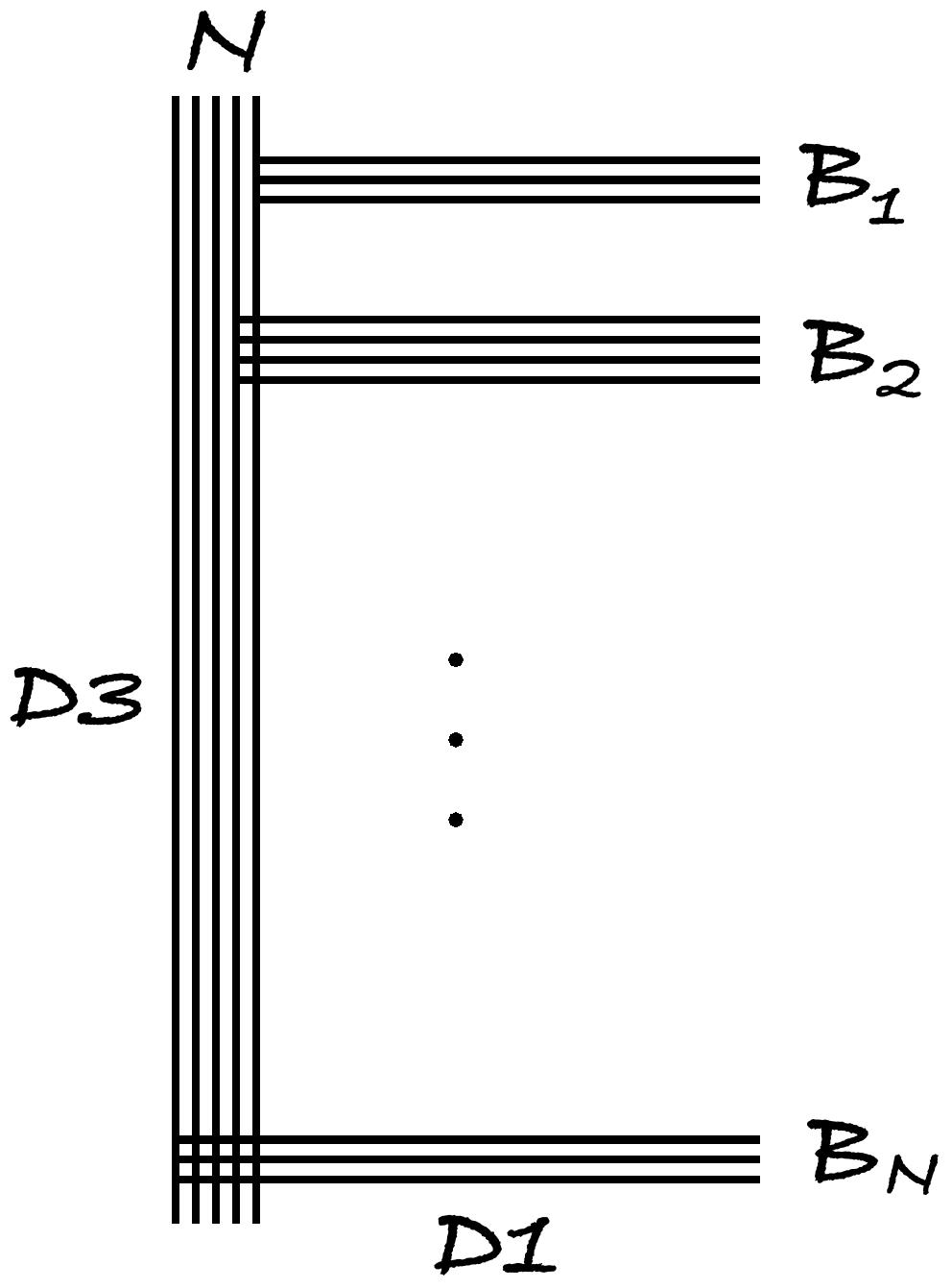}}
\caption{\label{Branes}\small A 't~Hooft line can be modeled by a stack of D1-branes ending on a stack of D3-branes. Because of the no-force condition the D1-branes are aligned in the transverse directions and are hence characterized by the same R-charge vector $n_i$, but can be split along the D3 world volume. An elementary object is thus a 't~Hooft loop with one unit of magnetic charge.}
\end{figure}

 The 't~Hooft loop operator is characterized by the magnetic charge vector $B$ and the R-charge orientation $n_i$, a unit vector in  six dimensions. The magnetic charges are embedded in the Cartan algebra of the gauge group:
\begin{equation}
 B=\mathop{\mathrm{diag}}(B_1,\ldots ,B_N),
\end{equation}
and obey the Dirac quantization condition, that requires all $B_i$'s to be integer. The brane configuration describing this setup is shown in fig.~\ref{Branes}.  The D1-branes can be split along the worldvolume of the D3's at no energy cost, in this sense the generic 't~Hooft line is a composite object consisting of elementary building blocks with unit magnetic charge. We will concentrate on the elementary 't~Hooft line and will set
\begin{equation}
 B=\mathop{\mathrm{diag}}(1,0,\ldots ,0), \label{B}
\end{equation}
that corresponds to a single D1-brane attached to one of the $D3$'s in the stack.

\begin{figure}[t]
\begin{center} 
\subfigure[]{
   \includegraphics[height=4cm] {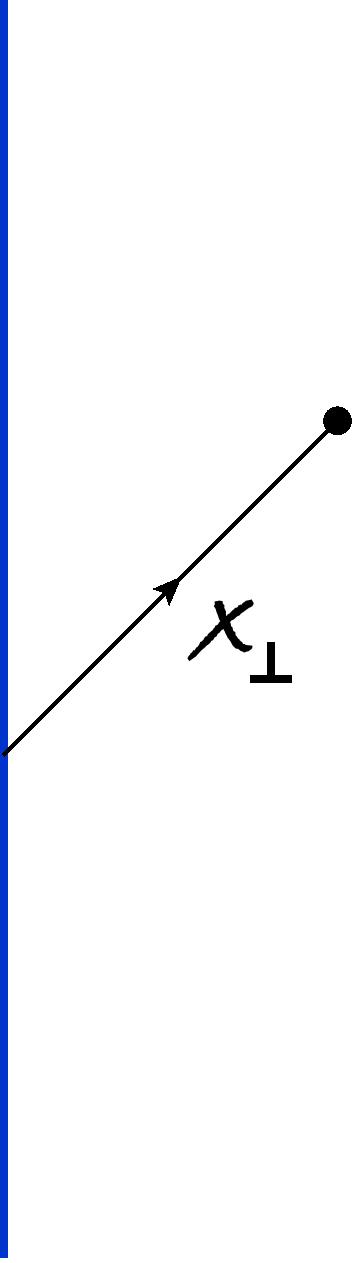}
   \label{straight}
 }
 \subfigure[]{
   \includegraphics[height=4cm] {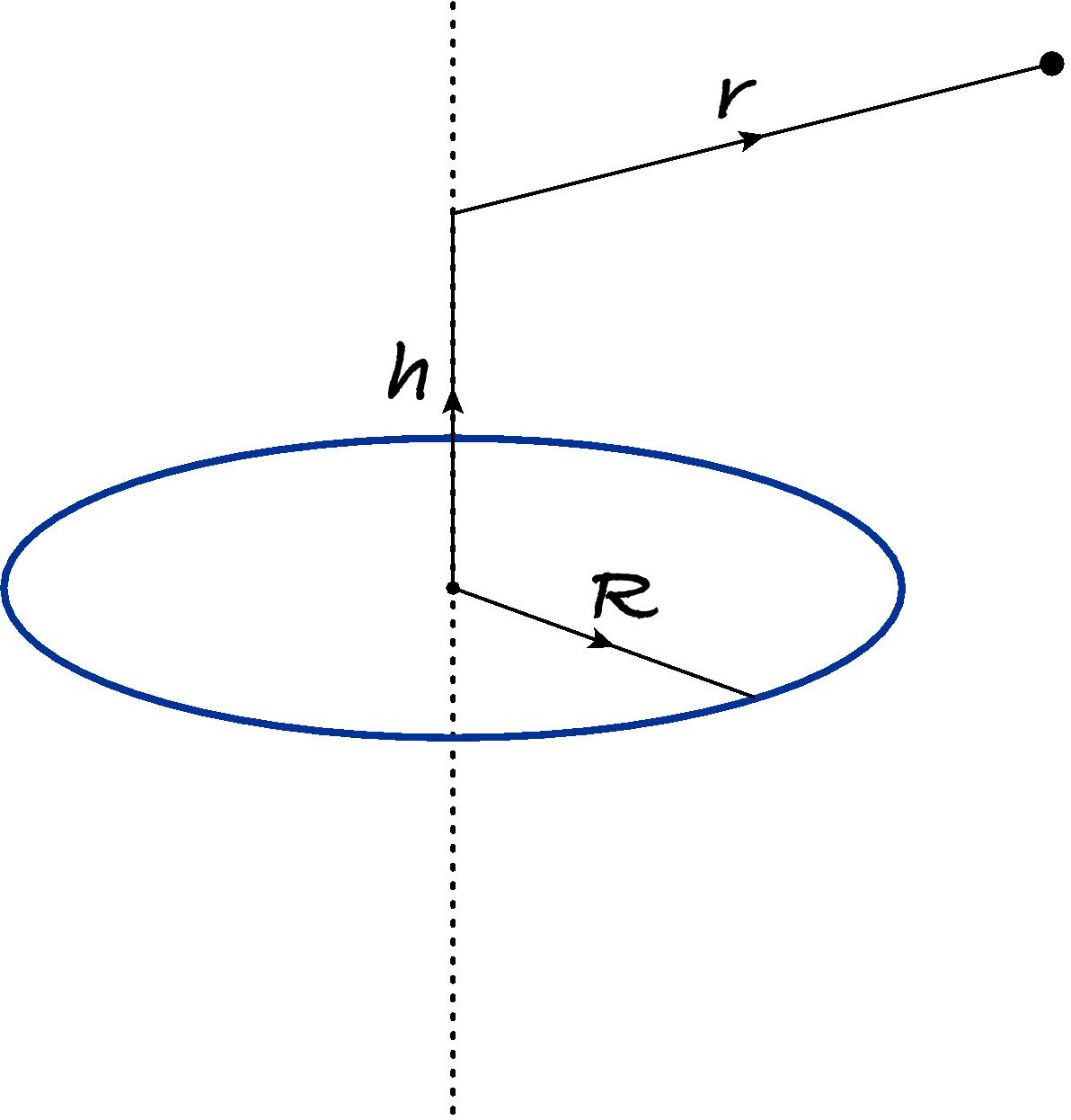}
   \label{general}
 }
  \subfigure[]{
   \includegraphics[height=4cm] {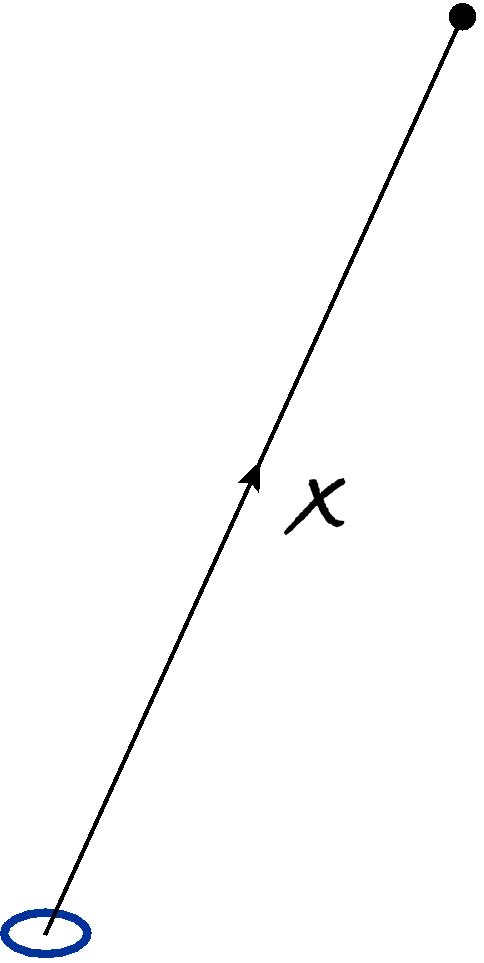}
   \label{far}
 }
\caption{\label{tot-fig}\small (a) A correlator between the 't~Hooft line and a local operator. (b) The correlator with the circular loop in general position. (c) The OPE limit thereof.}
\end{center}
\end{figure}

The straight 't~Hooft line in  $\mathcal{N}=4$ SYM preserves scale invariance along with
some supersymmetry\footnote{The subalgebra of $\mathfrak{psu}(2,2|4)$ preserved by the 't~Hooft line is $\mathfrak{osp}(4^*|4)$ \cite{Liendo:2016ymz}, the same as for the circular Wilson loop \cite{Bianchi:2002gz}.} and thus defines a supersymmetric defect CFT. Local operators can be confined to the defect or inserted in the bulk, for instance the correlator of the 't~Hooft line with a local operator defines a one-point bulk correlation function:
\begin{equation}
 \left\langle \mathcal{O}_a(x)\right\rangle_T\equiv \frac{\left\langle T(L)\mathcal{O}_a(x)\right\rangle}{\left\langle T(L)\right\rangle}\,.
\end{equation}
Since the only scale in the problem is the distance from the line to the operator insertion, fig.~\ref{straight},
the scale symmetry alone fixes the one-point function up to a single constant:
\begin{equation}\label{1pt}
 \left\langle \mathcal{O}_a(x)\right\rangle_T=\frac{\CC_a}{\left(2|x_\perp|\right)^{\Delta _a}}\,.
\end{equation}
The constant depends on the operator normalization but once normalization is fixed carries dynamical information on both the defect and operator. We fix the normalization ambiguity by demanding the standard form of the two-point function far from the defect:
\begin{equation}\label{field-Norm}
 \left\langle \mathcal{O}^\dagger _a(x)\mathcal{O}_b(0)\right\rangle=\frac{\delta _{ab}}{|x|^{2\Delta _a}}\,.
\end{equation}

An extra $2$ in (\ref{1pt}) is a rather standard convention, and is introduced for the following reason. An inversion map makes a circle from a line, fig.~\ref{general}. Applying inversion to (\ref{1pt}) and reading off the geometric data  one finds the correlator of a circular loop with  a local operator in general position:
\begin{equation}\label{genp}
\frac{ \left\langle T(C)\mathcal{O}_a(x)\right\rangle}{\left\langle T(C)\right\rangle}
 =\frac{\CC_aR^{\Delta _a}}{\left[h^2+(r-R)^2\right]^{\frac{\Delta _a}{2}}\left[h^2+(r+R)^2\right]^{\frac{\Delta _a}{2}}}\,.
\end{equation}
Consistency with (\ref{1pt}) can be checked by setting $r=R$ and sending $h\rightarrow 0$.

In the opposite limit, fig.~\ref{far},
\begin{equation}\label{large-x}
 \frac{ \left\langle T(C)\mathcal{O}_a(x)\right\rangle}{\left\langle T(C)\right\rangle}
 \stackrel{x\rightarrow \infty }{\simeq }\frac{\CC_aR^{\Delta _a}}{|x|^{2\Delta _a}}\,.
\end{equation}
The factor of two disappeared.
The formula shows  that $\CC_a$ is the OPE coefficient of the 't~Hooft loop expanded in the basis of local operators \cite{Shifman:1980ui}:
\begin{equation}
 T(C)=\left\langle T(C)\right\rangle\left(
 1+\sum_{a}^{}\CC_aR^{\Delta _a}\mathcal{O}_a(0)
 \right).
\end{equation}
The normalization is such that $\CC_{\mathbf{1}}=1$ and all of the OPE coefficients are dimensionless. Those latter will be the focus of further discussion.

\section{S-duality and protected operators \label{S-duality}}

The expectation value of the circular Wilson loop and its correlators with chiral primary operators admit a matrix model representation  \cite{Erickson:2000af,Drukker:2000rr,Semenoff:2001xp} that can be derived from supersymmetric localization on $S^4$ \cite{Pestun:2007rz}. An equivalent representation is through 2d Yang-Mills theory \cite{Giombi:2009ds}. Either way, the resulting matrix model is Gaussian and the correlators can be computed exactly, at any value of the gauge coupling and for any $N$ \cite{Drukker:2000rr,Okuyama:2006jc}. The correlators of the 't~Hooft loop follow from  S-duality which inverts the Yang-Mills coupling (not the 't~Hooft coupling that we use throughout the paper):
\begin{equation}
 \lambda \rightarrow \frac{16\pi ^2N^2}{\lambda }\,.
\end{equation}
The S-duality and the large-$N$ limit do not commute, so to reconstruct the 't~Hooft loop we really need to know the Wilson loop at any finite $N$, not just in the 't~Hooft limit.

\subsection{'t~Hooft loop expectation value}

The expectation value of the circular Wilson loop is known exactly at any value of the coupling \cite{Drukker:2000rr}:
\begin{equation}\label{Wvev}
 \left\langle W\right\rangle=\,{\rm e}\,^{\frac{\lambda }{8N}}L_{N-1}^1\left(-\frac{\lambda }{4N}\right),
\end{equation}
where $L_n^m(x)$ are the Laguerre polynomials. Applying S-duality we get for the expectation value of the circular 't~Hooft loop:
\begin{equation}\label{thl}
 \left\langle T\right\rangle=\,{\rm e}\,^{\frac{2\pi ^2N}{\lambda }}L_{N-1}^1\left(-\frac{4\pi ^2N}{\lambda }\right).
\end{equation}

The large-$N$ limit for the 't~Hooft and Wilson loops are quite different because the factors of $N$ are reversed by  S-duality. Yet, the 't~Hooft loop in the planar limit appears to be related to the multiply-wound Wilson loop in the limit when the number of windings scales with $N$  \cite{Drukker:2005kx}. The $k$-wound Wilson loop is given by (\ref{Wvev}) with $\lambda $ replaced by $\lambda k^2$. If we now take the simultaneous limit of $N\rightarrow \infty $, $k\rightarrow \infty $ with $\kappa =\lambda k/N$ fixed the problem becomes formally equivalent to computing the 't~Hooft loop in the conventional large-$N$ limit with $\kappa =4\pi /\sqrt{\lambda }$. 

The multiply-wound Wilson loop is known to exponentiate  \cite{Drukker:2005kx,Okuyama:2006jc,Hartnoll:2006is} and the expectation value of the 't~Hooft loop can be 
inferred from that result just by replacing $\kappa$ with $4\pi /\sqrt{\lambda }$:
\begin{eqnarray}\label{T-exp}
 \left\langle T\right\rangle&\stackrel{N\rightarrow \infty }{\simeq }&\,{\rm e}\,^{NF_T(\lambda )},
\\
\label{F_T}
F_T(\lambda )&=&\frac{2\pi }{\sqrt{\lambda }}\,\sqrt{1+\frac{\pi ^2}{\lambda }}
+2\ln\left(\sqrt{1+\frac{\pi ^2}{\lambda }}+\frac{\pi }{\sqrt{\lambda }}\right).
\end{eqnarray}
The next order in $1/N$ can be found in  \cite{Kawamoto:2008gp,Beccaria:2020ykg}.
Derivation of this result by a direct computation is very easy and is sketched below.

It follows from an integral representation of (\ref{thl}):
\begin{equation}
 \left\langle T\right\rangle=\frac{\sqrt{\lambda N}}{2\pi N!}\,\,{\rm e}\,^{-\frac{2\pi ^2N}{\lambda }}
 \int_{0}^{\infty }dt\,t^{N-\frac{1}{2}}\,{\rm e}\,^{-t}I_1\left(4\pi \sqrt{\frac{Nt}{\lambda }}\right).
\end{equation}
Changing variables $t=N\tau $, using the Stirling formula for $N!$ and taking the asymptotics of the Bessel function we get:
\begin{equation}\label{approx-T}
 \left\langle T\right\rangle\simeq 
 \frac{\lambda ^{\frac{3}{4}}}{8\pi ^{\frac{5}{2}}}
 \int_{0}^{\infty }d\tau \,
 \tau ^{-\frac{3}{4}}
 \,{\rm e}\,^{-NS(\tau )},
\end{equation}
where
\begin{equation}\label{S(tau)}
 S(\tau )=\tau -\ln\tau -4\pi \sqrt{\frac{\tau }{\lambda }}+\frac{2\pi ^2}{\lambda }-1.
\end{equation}
The main contribution to the integral comes from the saddle point at
\begin{equation}\label{taustar}
 \sqrt{\tau_* }=\sqrt{1+\frac{\pi ^2}{\lambda }}+\frac{\pi }{\sqrt{\lambda }}\,,
\end{equation}
and in the leading approximation we get (\ref{T-exp}), (\ref{F_T}). 

The localization prediction is consistent with the semiclassical nature of the 't~Hooft loop. At weak coupling:
\begin{equation}
 F_T\stackrel{\lambda \rightarrow 0}{\simeq }
 \frac{2\pi ^2}{\lambda }+\ln\frac{4\pi ^2}{\lambda }+\ldots 
\end{equation}
The first term is just the Yang-Mills action evaluated on the field of the monopole \cite{Drukker:2005kx,Gomis:2009ir}, while the log-correction comes from the zero modes and the gauge group volume  \cite{Drukker:2005kx,Gomis:2009ir}.

On the strong-coupling side:
\begin{equation}
 F_T\stackrel{\lambda \rightarrow \infty }{\simeq }\frac{4\pi }{\sqrt{\lambda }}+\ldots 
\end{equation}
This can be compared to the action of the hemi-spherical D1-brane ending on the circular loop at the boundary of $AdS_5$. Indeed, the D1-brane tension is\footnote{This is the standard textbook formula \cite{polchinski1998string} re-written in the AdS units, whereupon the string tension is defined by the dimensionless ratio: $\alpha '/R^2=1/\sqrt{\lambda }$, and the string coupling is traded for the gauge coupling: $g_s=4\pi \lambda /N$.}
\begin{equation}\label{tension}
 T_{D1}=\frac{2N}{\sqrt{\lambda }}\,,
\end{equation}
and the regularized volume of the minimal surface equals  $A=-2\pi $ \cite{Berenstein:1998ij,Drukker:1999zq}. Thus,
$$
 \left\langle T\right\rangle\simeq \,{\rm e}\,^{-T_{D1}A}=\,{\rm e}\,^{\frac{4\pi N}{\sqrt{\lambda }}}\,,
$$
in agreement with the result obtained by localization.

\subsection{Correlators with chiral primaries}

Consider chiral primary operators
\begin{equation}\label{O_L}
 \mathcal{O}_L=\frac{1}{\sqrt{L}}\,\left(\frac{4\pi ^2}{\lambda }\right)^{\frac{L}{2}}\mathop{\mathrm{tr}}Z^L.
\end{equation}
The overall factor makes them unit-normalized. To make contact with localization we need to align the R-charge of the operator to that of the 't~Hooft loop. For that matter,  we assign the orientation vector $n_i=(1,0,\ldots ,0)$ to the 't~Hooft loop and define the operator's polarization by $Z=\Phi _1+i\Phi _2$.

 For the Wilson loop with the same R-charge assignment the one-point function is known exactly \cite{Okuyama:2006jc}:
\begin{equation}
 \left\langle W \mathcal{O}_L(x)\right\rangle=\frac{1}{\sqrt{L}}\,
 \left(\frac{\sqrt{\lambda}}{2NR}\right)^{L}
 \,{\rm e}\,^{\frac{\lambda }{8N}}
 \sum_{k=1}^{L}L_{N-k}^L\left(-\frac{\lambda }{4N}\right).
\end{equation}
This formula is literally correct if the operator is inserted in the centre of the circle, otherwise  $R$ should be replaced by the appropriately defined conformal distance  from (\ref{genp}). 

Applying S-duality we get for the correlator with the 't~Hooft loop:
\begin{equation}\label{TOL}
 \left\langle T\mathcal{O}_L(x)\right\rangle
 =\frac{1}{\sqrt{L}}\,\left(\frac{2\pi }{\sqrt{\lambda}\, R}\right)^{L}
 \,{\rm e}\,^{\frac{2\pi ^2N}{\lambda }}\sum_{k=1}^{L}L_{N-k}^L\left(-\frac{4\pi ^2N}{\lambda }\right).
\end{equation}
 The large-$N$ limit can again be extracted from the integral representation:
\begin{equation}
 \left\langle T\mathcal{O}_L(x)\right\rangle=
 \frac{1}{\sqrt{\lambda }\,N^{\frac{L}{2}}R^{L}}\,
  \,{\rm e}\,^{-\frac{2\pi ^2N}{\lambda }}
  \int_{0}^{\infty }dt\,\,{\rm e}\,^{-t}
  \sum_{k=1}^{L}\frac{t^{N-k+\frac{L}{2}}}{(N-k)!}\,I_L\left(4\pi \sqrt{\frac{tN}{\lambda }}\right).
\end{equation}

Changing variables to $\tau =tN$ and taking the large-$N$ limit of all the prefactors we get:
\begin{equation}
  \left\langle T\mathcal{O}_L(x)\right\rangle\simeq 
  \frac{\lambda ^{\frac{1}{4}}}{4\pi ^{\frac{3}{2}}\sqrt{L }\,R^{L}}
  \int_{0}^{\infty }d\tau \,\sum_{k=1}^{L}\tau ^{\frac{L}{2}-k-\frac{1}{4}}
  \,{\rm e}\,^{-NS(\tau )},
\end{equation}
with $S(\tau )$ given by (\ref{S(tau)}). Evaluating the integral in the saddle-point approximation and normalizing by (\ref{approx-T}) we get:
\begin{equation}
 \left\langle \mathcal{O}_a(x)\right\rangle_T
 =\frac{2\pi }{\sqrt{\lambda L}\,R^{L}}\sum_{k=1}^{L}\tau _*^{\frac{L}{2}-k+\frac{1}{2}}=\frac{2\pi}{\sqrt{\lambda L}\,R^{L}}\,\,
 \frac{\tau _*^{\frac{L}{2}}-\tau _*^{-\frac{L}{2}}}{\tau _*^{\frac{1}{2}}-\tau _*^{-\frac{1}{2}}},
\end{equation}
with $\tau _*$ from (\ref{taustar}). Comparing to (\ref{large-x}) we read off the OPE coefficient:
\begin{equation}\label{CPO-OPE}
 \CC_L=\frac{1}{\sqrt{L}}
 \left[
 \left(\sqrt{1+\frac{\pi ^2}{\lambda }}+\frac{\pi }{\sqrt{\lambda }}\right)^L
 - \left(\sqrt{1+\frac{\pi ^2}{\lambda }}-\frac{\pi }{\sqrt{\lambda }}\right)^L
 \right].
\end{equation}

Foreseeing an integrability description we can interpret the first term as an asymptotic answer, to be dressed by Bethe-ansatz structures for non-protected operators, while the second term is naturally interpreted as a wrapping correction~\cite{Ambjorn:2005wa}. 
The two terms are indeed separated by $L$ orders of perturbation theory as appropriate for diagrams "wrapping the cylinder".

We can make the argument sharper by introducing the Zhukowski variable
\begin{equation}
 x(u)+\frac{1}{x(u)}=\frac{4\pi u}{\sqrt{\lambda }}\,,
\end{equation}
and denoting
\begin{equation}
 x\left(\frac{ai}{2}\right)\equiv ix_a .
\end{equation}
Then (\ref{CPO-OPE}) can be concisely written as
\begin{equation}\label{1pt-Zhukowski}
 \CC_L=\frac{1}{\sqrt{L}}\left(x_1^L-\frac{1}{x_1^L}\right).
\end{equation}
Structurally, this is very similar to the one-point function in the D3-D5 dCFT \cite{Komatsu:2020sup} but is much simpler. In that latter case the asymptotic part is a sum of $k$ terms ($k$ is a quantum number of the defect)  of the form  $x_a^L$ with $a$ running between $1-k$ and $k-1$, and the wrapping corrections are a series in $1/x_a^L$. Each of the asymptotic terms corresponds to a particular bound state formed by $a$ elementary magnons with the D5-brane. There is only one asymptotic term here and we expect that the string modes do not form bound states, so the scattering theory for the 't~Hooft loop must be way simpler.

Expanding the one-point function at weak coupling, we find:
\begin{equation}
 \CC_L\simeq \frac{1}{\sqrt{L }}\left(\frac{4\pi ^2}{\lambda }\right)^{\frac{L}{2}} 
 \left(1+\frac{g_{\footnotesize YM}^2 (N-1)}{4\pi^2} L +\ldots
 \right). \label{prediction}
\end{equation}
In the 't~Hooft limit this formula is just Taylor series of (\ref{CPO-OPE}), but written that way remains valid at any $N$, as follows from (\ref{TOL}), (\ref{thl}) and the asymptotic form of the Laguerre polynomials:
\begin{equation}
 L_n^s(x)=\frac{x^n}{n!}\left[1+\frac{n(n+s)}{x}+\ldots \right].
\end{equation}

\begin{figure}[t]
\begin{center} 
\subfigure[]{
   \includegraphics[height=2.5cm] {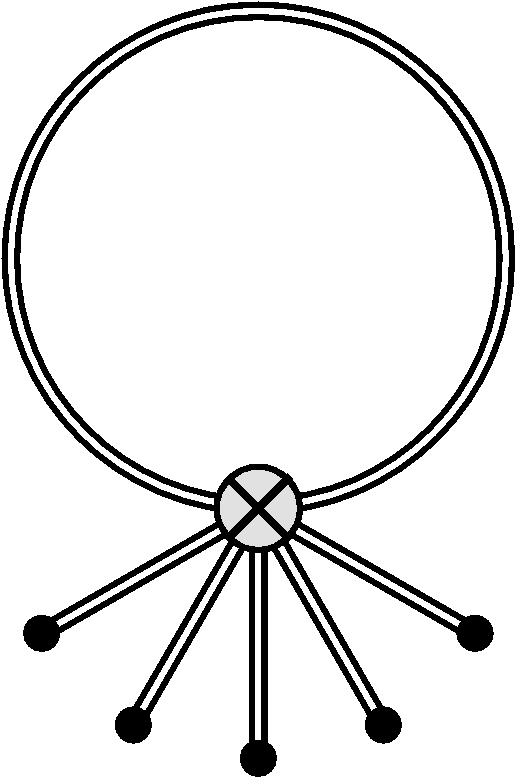}
   \label{planar-fig}
 }
 \subfigure[]{
   \includegraphics[height=2.5cm] {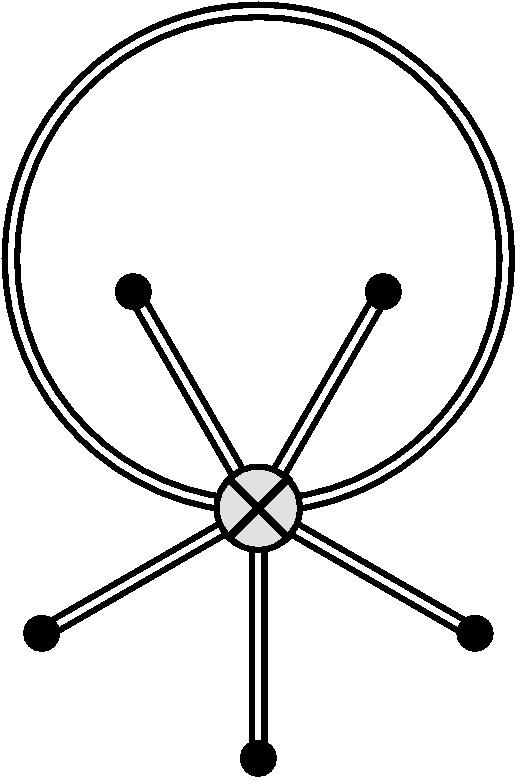}
   \label{non-planar-fig}
 }
\caption{\label{tot-diag-fig}\small The one-loop correction to the one-point function. The dots denote the classical field (\ref{F_cl}), (\ref{Phi_cl}). The loop propagator is the Green's function in this background.}
\end{center}
\end{figure}

The leading order is consistent with substituting the classical field (\ref{Phi_cl})  into (\ref{O_L})\footnote{It is important to recall that the classical field (\ref{Phi_cl}) describes an infinite 't~Hooft line, fig.~\ref{straight}, correlator which has an extra factor of $1/2^{\Delta }$ compared to the circular model, see (\ref{1pt}).}. The next order can likewise be
reproduced by a quantum field theoretical calculation as we will demonstrate in section~\ref{QFT} by explicitly calculating the planar diagram in fig.~\ref{planar-fig} and showing that the non-planar corrections in fig.~\ref{non-planar-fig} vanish.

Planar corrections at the next  $L$ orders come entirely from the first term in (\ref{CPO-OPE}). The simplicity of the answer suggests many cancellations and points to simple combinatorics of the higher-loop corrections. 
We can also compare the exact OPE coefficient with supergravity and string theory at strong coupling. This analysis is
relegated to appendix~\ref{sugra}.

\section{Quantizing 't Hooft lines \label{QFT}}

In this section we will set up the program for doing perturbative calculations in the monopole background, and as a
first application we reproduce the one-loop result predicted by S-duality~(\ref{prediction}). Various aspects of the quantization
of 't Hooft lines have been discussed in~\cite{Gomis:2009ir,Gomis:2009xg}. The problem is similar to quantization of  monopole operators in 3d \cite{Borokhov:2002ib,Borokhov:2002cg}, which has been discussed in the gauge theory with sixteen supercharges (the dimensional reduction of $\mathcal{N}=4$ SYM) \cite{Grignani:2007xz}. Unfortunately we cannot use these results directly because the theory is not conformal in 3d and the mixing pattern of various field components is different as a result.

\subsection{Expanding around the monopole background}
We start from the ${\cal N}=4$ SYM action in the following form
\begin{equation}
S=\frac{2}{g_{\footnotesize YM}^2} \int d^4 x \,\mbox{tr} \, \left[ -\frac{1}{4} F_{\mu\nu} F^{\mu \nu} -
\frac{1}{2} D_{\mu} \varphi_i D^{\mu} \varphi_i+ \frac{1}{4} [\varphi_i,\varphi_j][\varphi_i,\varphi_j] + \mbox{fermions}\right],
\end{equation}
where
\begin{equation}
F_{\mu\nu}=\partial_\mu A_\nu-\partial_\nu A_\mu -
i[A_\mu,A_\nu], \hspace{0.5cm} D_\mu \varphi_i= \partial_{\mu} \varphi_i -i[A_\mu,\varphi_i], 
\end{equation}
with $\mu,\nu=0,1,2,3$ and $i,j=1,2,\ldots,6$,
and we wish to expand around the classical field configuration given by \eqref{F_cl}, \eqref{Phi_cl} and \eqref{B}. 
Inserting
\begin{equation}
\varphi_i=\varphi_i^{\text cl}+ \tilde{\varphi}_i, \hspace{0.5cm} A_{\mu}=A_{\mu}^{\text cl}+\tilde{A}_\mu,
\end{equation}
and adding a gauge fixing term 
\begin{equation}
S_{gf}=\frac{2}{g_{\footnotesize YM}^2} \int d^4 x\, \mbox{tr}\, \left[- \frac{1}{2} \left(\partial_{\mu} \tilde{A}^\mu +i
[\tilde{A}^\mu,A_{\mu}^{{\text cl}}]+
i[\tilde{\varphi}_i,\varphi_i^{\text cl}] \right)^2\right],
\end{equation}
we get for the resulting action\footnote{Strictly speaking we should consider the full ${\cal N}=4$ SYM
and perform the gauge fixing by adding a BRST exact term as in~\cite{Buhl-Mortensen:2016pxs,Buhl-Mortensen:2016jqo}. }
\begin{eqnarray}
S&=&\frac{2}{g_{\footnotesize YM}^2} \int d^4 x \,\mbox{tr} \, \left \{
-\frac{1}{2} \bar{D}_{\nu} A_\mu \bar{D}^{\nu} A^{\mu} -\frac{1}{2} \bar{D}_\nu \varphi_i \bar{D}^{\nu} \varphi_i \right .\nonumber \\
&& \hspace{1.0cm}+i\bar{F}_{\mu \nu}  [A^{\mu},A^{\nu}]  +2i \,\partial_{\mu} \varphi_i^{\text cl} [A^\mu,\varphi_i]
\nonumber\\ 
&&\ \left.  \hspace{1.0cm}+\frac{1}{2}[A_\mu, \varphi_i^{\text cl}][A^\mu,\varphi_i^{\text cl}] +\frac{1}{2}[\varphi_j, \varphi_i^{\text cl}][\varphi_j,\varphi_i^{\text cl}] \right\}, \label{Sgf}
\end{eqnarray}
where we have omitted the tildes on the quantum fields and where
\begin{equation}
\bar{D}_{\nu}=\partial_{\nu} -i[ A_{\nu}^{\text cl},\,\cdot\,].
\end{equation}
The symmetry breaking pattern of the classical fields implies the following block decomposition of the quantum fields
\begin{equation}\label{Decomposition}
\begin{array} {ll}
\hphantom{A_\mu ,\varphi _i,\Psi  = }~\!1\,\,\,~~~N-1 &  \\[1pt]
   A_\mu ,\varphi _i= \left[
 \begin{array}{c:ccc}
 \kkb & \knb & \knb & \knb \\
 \hdashline
  \nkb & \nnb  & \nnb & \nnb \\
   \nkb & \nnb & \nnb & \nnb  \\
    \nkb & \nnb & \nnb & \nnb \\
 \end{array}
 \right]
 \begin{array}{c}
 1\\
   \\
  N -1 \\
    \\
 \end{array}
  \end{array}
\end{equation}
Here the action for the field components of type \ae\  as well as for the component \aa\ are unaffected by the presence of the defect whereas new quadratic terms appear for the fields of type \o.

We  now choose $n_i=(1,0,0,0,0,0)$,  and clearly we can set $A_0^{\text cl}=0$, cf.\ eqns.~\eqref{Phi_cl}, \eqref{F_cl}.
Next, we can replace $\varphi_i^{\text cl}$ with $\varphi_1^{\text cl}$  in the expression~\eqref{Sgf} for the action, and we observe that the fields split into two groups, namely the simple fields
$\{A_0, \varphi_2,\varphi_3,\varphi_4, \varphi_5, \varphi_6\}$ for which the quadratic terms are diagonal in flavour and 
the complicated fields $\{\varphi_1,A_1,A_2,A_3\}$ which mix between each other.  A convenient gauge choice for the remaining
components of $A^{\text cl}$~\cite{Dirac:1931kp} is\footnote{The gauge connection is $A=B(1-\cos\theta )d\phi/2$, whose curvature is indeed the volume form of $S^2$: $F=Bd\phi \wedge d\theta /2$. The factor of $r\sin\theta $ arises from transformation to the local frame of the spherical coordinate system.}
\begin{equation}
A_\phi^{\text cl}=\frac{B}{2r}\frac{1-\cos \theta}{ \sin \theta}, \hspace{0.5cm} A_{r}^{\text cl}=A_{\theta}^{\text cl}=0,
\end{equation} 
or expressed in Cartesian coordinates
\begin{equation}
A_x^{\text cl}=-\frac{By}{2r(r+z)}, \hspace{0.5cm} A_{y}^{\text cl}=\frac{Bx}{2r(r+z)}, \hspace{0.5cm}
A_{z}^{\text cl}=0.
\end{equation}

\subsection{Dirac's quantum mechanical monopole problem \label{QM}}

\begin{figure}[t]
\begin{center} 
\subfigure[]{
   \includegraphics[height=3.0cm] {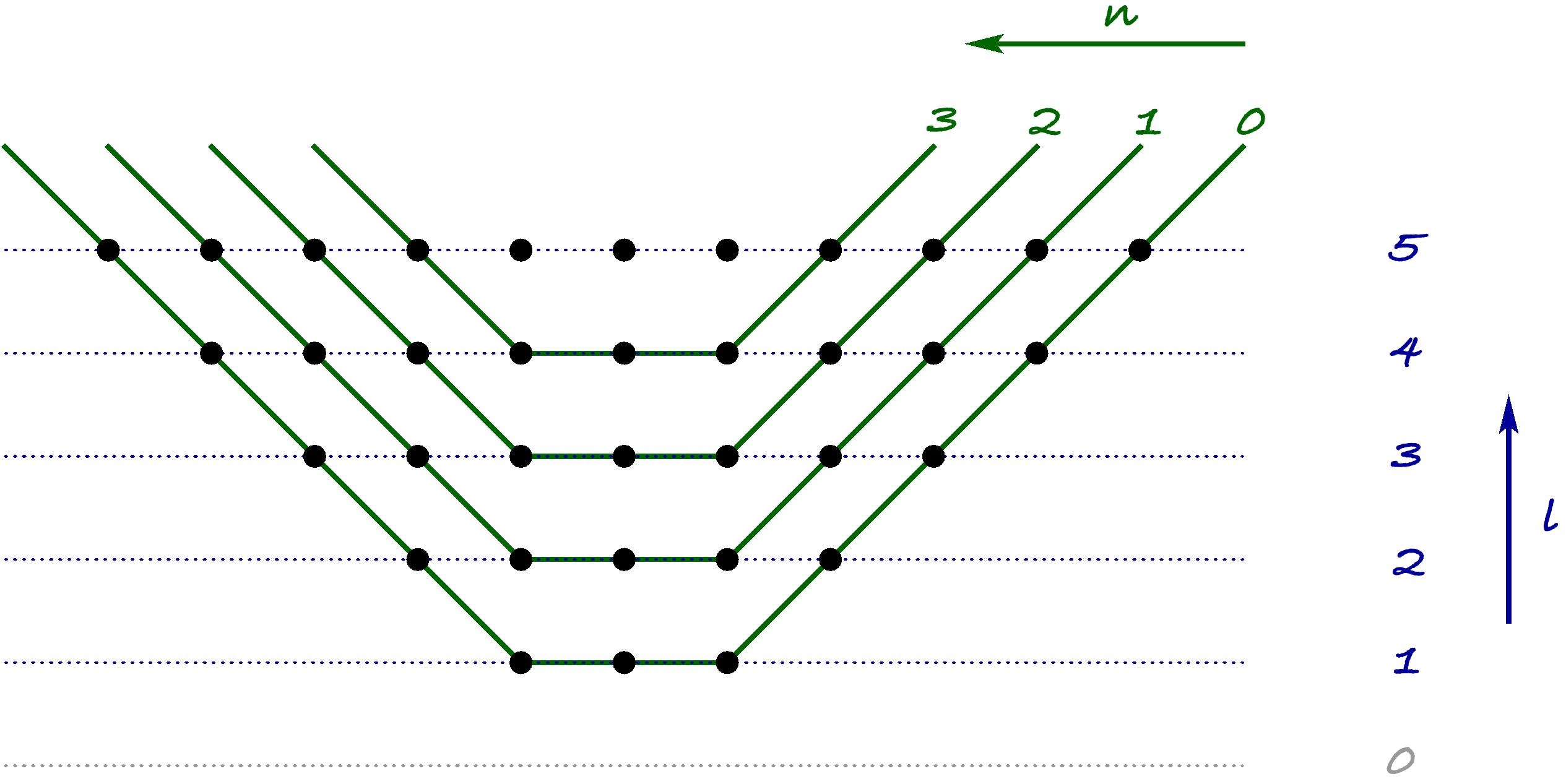}
   \label{spec}
 }
\subfigure[]{
   \includegraphics[height=2.3cm] {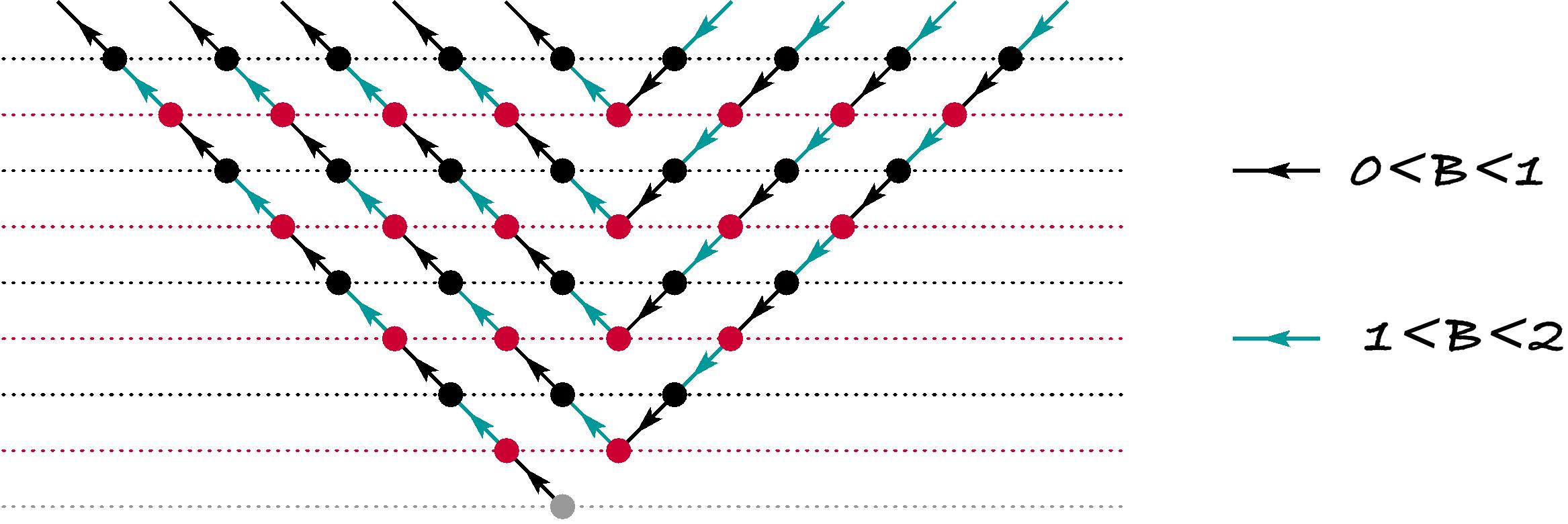}
   \label{flow}
 }
\caption{\label{tot-spectrum}\small (a) The spectrum, at $B=2$. (b) The spectral flow from $B=0$ to $B=2$. The Dirac quantization condition corresponds to crossing (half-)integer-$l$ lines when the spectrum recombines into the usual $SU(2)$ multiplets. The zero angular momentum state disappears in the process.}
\end{center}
\end{figure}

Determining the propagators for the fields without  flavour mixing 
amounts to solving Dirac's original quantum mechanical monopole problem~\cite{Dirac:1931kp}.
The starting point is the 
quadratic term in the
Lagrangian density for the \o-components of the simple fields which reads
\begin{eqnarray}
{\cal L}_s&=& 
\varphi_{1j} \mbox{{\Large (}} \partial^2+2 i A_\nu^{\text cl}\,\partial_\nu-(A_\nu)^{\text cl} (A^{\nu})^{\text cl}-(\varphi_1^{\text cl})^2 \mbox{{\Large)}} \varphi_{j1}, \\
&=&
\varphi_{j1}^{\dagger}
\left( \partial_t^2
+(\partial^k+iA_{\text cl}^k) (\partial_k+iA^{\text cl}_k)-\frac{B}{4 r^2}
\right) \varphi_{j1},
 \hspace{0.5cm} j\neq 1, \label{Lsimple}
\end{eqnarray}
where it is understood that the classical fields are replaced by their non-vanishing 11-component and where we have introduced the notation
\begin{equation}
\varphi_{1j}^{\dagger}=(\varphi^\dagger)_{1j}=\varphi_{j1}.
\end{equation}
Furthermore, the index $k$ now runs over spatial values only. 
In order to determine the propagators of the simple fields we shall start by determining the spectrum of the operator
\begin{equation}
\hat{H}=- (\partial^k+iA_{\text cl}^k) (\partial_k+iA^{\text cl}_k)+\frac{B}{4 r^2},
\end{equation}
i.e. solving the eigenvalue problem
\begin{equation}
\hat{H}\Psi=E\Psi. \label{eigeneqn}
\end{equation}
Apart from the last purely radial term, this problem is identical to Dirac's original quantum mechanical monopole problem~\cite{Dirac:1931kp} whose solution we now briefly review following ~\cite{Tamm:1931dda,Fierz:1944,Wu:1976qk}, see also~\cite{Roy:1983gu}.
It is convenient to work in spherical coordinates where the Hamiltonian reads
\begin{equation}
\hat{H}=-\left(\nabla^2 + \frac{i\, B}{r^2 (1+\cos \theta)}\partial_\phi-\frac{B^2}{4 r^2}\frac{1-\cos\theta}{1+\cos\theta}
-\frac{B^2}{4 r^2}\right),
 \label{Hamiltonian}
\end{equation}
with $\nabla^2$  the 3d Laplacian.
 Following~\cite{Fierz:1944} we define
\begin{equation}
L_{\pm}= L_x\pm iL_y= e^{\pm i\phi}\left[\pm \frac{\partial}{\partial \theta} +
i\cot \theta \frac{\partial}{\partial \phi}+\frac{B}{2} \frac{\sin\theta}{1+\cos \theta}\right],
\end{equation}
\begin{equation}
L_z=-i\frac{\partial}{\partial \phi} +\frac{B}{2}.
\end{equation}
Then we have 
\begin{equation}
\hat{H}= p_r^2 +\frac{L^2(\theta,\phi)}{r^2}, \hspace{0.7cm} \mbox{where}\hspace{0.7cm}p_r=\frac{-i}{r}\frac{\partial}{\partial r}  r,
\end{equation}
and 
\begin{equation}
L^2(\theta,\phi)=-\left[
 \frac{1}{\sin \theta}\,\frac{\partial}{\partial \theta}\,\sin\theta \,\frac{\partial }{\partial \theta} +
 \frac{1}{\sin^2 \theta}  \frac{\partial^2}{\partial \phi^2}+\frac{iB}{1+\cos\theta}\partial_\phi-\frac{B^2}{2} \frac{1}{1+\cos\theta}\right],
\label{L2def}
\end{equation}
Now, it holds that 
\begin{equation}
\left[\vec{L},H\right]=0,
\end{equation}
and the operators $L_i$ fulfill the familiar $SU(2)$ commutation relations
\begin{equation}
\left[ L_i,L_j\right]=i \epsilon_{ijk} L_k.
\end{equation}
The solution to the eigenvalue problem is found by assuming that the wave function
factorizes as follows
\begin{equation}
\Psi= \rho(r) \,W(\theta, \phi),
\end{equation}
which implies that~\eqref{eigeneqn} becomes equivalent to the following two equations
\begin{equation}
 \frac{\partial^2 \rho}{\partial r^2}  +\frac{2}{r} \frac{\partial \rho}{\partial r} -\frac{\lambda}{r^2}\rho = -E \rho,   \label{radial} 
\end{equation}
\begin{equation}
L^2(\theta,\phi) W(\theta,\phi)= \lambda W(\theta,\phi) \label{angular1}.
\end{equation}
We can  immediately conclude that the angular wave functions will constitute highest weight representations of $SU(2)$.
Introducing the further factorization 
\begin{equation}
W(\theta, \phi)= U(\theta) e^{im\phi}, \hspace{0.5cm} m=0, \pm 1,\pm 2, \ldots,
\end{equation}
where $m$ has to be integer for the wave function to be single valued, we find
\begin{equation}
L_z W(\theta,\phi)= \left(m+\frac{B}{2} \right) W(\theta,\phi) \equiv \overline{m} \,W(\theta,\phi), \label{Jzvalue}
\end{equation}
and the angular wave equation takes the form
 \begin{equation}
- \frac{1}{\sin \theta}\,\frac{\partial}{\partial \theta}\,\sin\theta \,\frac{\partial U}{\partial \theta}+ \left(\frac{m^2}{\sin^2 \theta} +\frac{mB}{1+\cos \theta}+\frac{B^2}{2}\frac{1}{1+\cos\theta}\,\right) U(\theta) =\lambda U(\theta). \label{angular}
\end{equation}
From SU(2) representation theory we know that the possible values of $\lambda$ are 
\begin{equation}
\lambda=l(l+1),
\end{equation}
where $l$ is a non-negative integer, when $B=0$. In the presence of the monopole, we encounter the phenomenon of spectral flow, i.e.\ the fact that the energy levels re-organize themselves as $B$ is varied.  In figure~\ref{tot-spectrum} we illustrate the spectral flow
between $B=0$ and $B=2$. For the case, we are interested in, i.e. $B=1$, we see that the possible 
 values of $l$ have shifted from integer values at  $B=0$ to half-integer values at $B=1$, cf.\ equation~\eqref{Jzvalue}.
Introducing the variable
\begin{equation}
x=\cos \theta,
\end{equation}
equation~\eqref{angular} can be written as
\begin{equation}
\left[(1-x^2) \frac{\partial^2}{\partial x^2} -2 x \frac{\partial }{\partial x} -
\frac{1}{2(1-x)}\left (\overline{m}-\frac{B}{2}\right)^2 - \frac{1}{2(1+x)} \left(\overline{m}+\frac{B}{2}\right)^2 \right] \,U=-\lambda U.
\label{angularx}
\end{equation}
Performing the following rescaling
\begin{equation}
U(x)=(1-x)^{\alpha/2} \,(1+x)^{\beta/2} \,\widetilde{U}(x),
\end{equation}
with
\begin{equation}
\alpha=\left |\,\overline{m}+\frac{B}{2}\right|, \hspace{0.5cm}\beta=\left |\,\overline{m}-\frac{B}{2}\right|,
\end{equation}
the equation~\eqref{angularx} turns into the following equation for $\widetilde{U}$
\begin{equation}
(1-x^2) \widetilde{U}'' +(\beta-\alpha-x(2+\alpha+\beta))\widetilde{U}'+\left(l(l+1)+\frac{1}{4}(\alpha+\beta)(2+\alpha+\beta\right)\widetilde{U}=0, \label{angularalpha}
\end{equation}
where the primes denote differentiation with respect to $x$. We recognize equation~\eqref{angularalpha} as the defining
equation for the Jacobi polynomials, i.e.
\begin{equation}
\widetilde{U}(x) ={\cal N} P_n^{(\alpha,\beta)} (x),
\end{equation}
with ${\cal N}$ a normalization constant, provided
\begin{equation}
\left(l(l+1)+\frac{1}{4}(\alpha+\beta)(2+\alpha+\beta\right)=n(n+\alpha+\beta+1), \hspace{0.5cm} n=0,1,2,\ldots, 
\end{equation}
i.e.
\begin{equation}
l=n+\frac{1}{2}(\alpha+\beta)= n+\frac{1}{2}\left |\,\overline{m}+\frac{B}{2}\right|+\frac{1}{2}\left |\,\overline{m}-\frac{B}{2}\right|,
\hspace{0.5cm} n=0,1,2,\ldots.
\end{equation}
More specifically the eigenfunctions of our angular problem read
\begin{eqnarray}
\lefteqn{
W(\theta,\phi)=Y^{(q)}_{l\bar{m}}(\theta,\phi)} \\
&=&C_{l\overline{m}}\, (1-\cos\theta)^{\alpha/2}(1+\cos\theta)^{\beta/2} P_{l-\alpha/2-\beta/2}^{(\alpha,\beta)} (\cos \theta) \,e^{i (\overline{m}-B/2)\phi},
\end{eqnarray}
where $C_{l\overline{m}}$ is a normalization constant and where the superscript $q$ denotes the monopole charge
\begin{equation}
q=\frac{B}{2}.
\end{equation}
These are the so-called monopole spherical harmonics and 
for the case of interest in this paper, i.e.\  $B=1$, 
\begin{equation}
l= \frac{1}{2}, \frac{3}{2},\ldots, \hspace{0.5cm} \overline{m}=-l,-l+1,\ldots,l-1,l,
\end{equation}
exposing again the phenomenon of spectral flow.

 The normalization constant can be determined explicitly and reads
\begin{equation}
C_{l \overline{m}}=\left(\frac{(2 l+1)  \left(-\frac{\alpha }{2}-\frac{\beta }{2}+l\right)! \,
   \Gamma \left(l+\frac{\alpha }{2}+\frac{\beta }{2}+1\right)}
   {2^{+\alpha +\beta +2}\,\pi\,\Gamma
   \left(l+\frac{\alpha }{2}-\frac{\beta }{2}+1\right)\Gamma \left(l-\frac{\alpha
   }{2}+\frac{\beta }{2}+1\right)}\right)^{1/2}.
\end{equation}

Returning to the radial equation, we insert $\lambda=l(l+1)$ and rescale $\rho(r)=r^{-1/2} R(r)$ which leads to  Bessel's
equation
\begin{equation}
\frac{d^2 R(r)}{d r^2}+\frac{1}{r} \frac{dR(r)}{dr} +\left(E-\frac{(l+1/2)^2}{r^2}\right)R(r)=0.
\end{equation}
 Picking the solution which makes $\rho(r)$ regular at $r=0$ gives 
 \begin{equation}
 \rho(r)=(\sqrt{E}r)^{-1/2}J_{\nu}(\sqrt{E}r),
 \end{equation}
 with
 \begin{equation}
 \nu=l+\frac{1}{2}. \label{nu}
 \end{equation}
 Having completed the quantization of the monopole problem we can construct the propagators of the simple fields via spectral decomposition, cf.\ section~\ref{Propagators}

 \subsection{An $S=1$ particle, a scalar and the monopole}
 
 Determining the propagators for the fields with flavour mixing amounts to determining the spectrum of a spin one particle 
 interacting with a scalar in the monopole background.
 The problem involving the spin one particle alone was analyzed in 
 \cite{Olsen:1990jm,Weinberg:1993sg} (we will follow \cite{Olsen:1990jm}). The starting point is 
the quadratic part of the Lagrangian density for the $\o$ components of the fields with mixing, i.e. $\varphi_1\equiv \varphi, A_1,A_2, A_3$, which reads
 \begin{eqnarray}
 {\cal L}_{m}&=&\varphi_{j1}^{\dagger}
\left( \partial_t^2
+(\partial^k+iA_{\text cl}^k) (\partial_k+iA^{\text cl}_k)-\frac{B}{4 r^2}
\right) \varphi_{j1} \\
&+ &  \vec{A}_{j1}^{\dagger}
\left( \partial_t^2
+(\nabla^k+iA_{\text cl}^k) (\nabla_k+iA^{\text cl}_k)-\frac{B}{4 r^2}\right )\vec{A}_{j1}\\
&+& i \varphi_{j1}^{\dagger} \left (\frac{B}{r^2} \,{\bf \hat{r}}\cdot \vec{A}_{j1}\right) -i \left( \vec{A}_{j1}^{\dagger} \cdot \frac{B}{r^2} \,{\bf \hat{r}}\right) {\varphi}_{j1} \\
&-&  \, i\vec{A}_{j1}^{\dagger} \cdot \left( \frac{B}{r^2}\,{\bf{\hat{r}}} \times \vec{A}_{j1}\right), 
 \hspace{0.5cm} j\neq 1,  \label{Lmixing}
 \end{eqnarray}
 where we have underlined the fact that the A-field constitutes a vector field. As for the simple fields we will start by diagonalizing the time-independent part of the problem. From section~\ref{QM} we already know how to diagonalize the term
 in the first line above making use of monopole spherical harmonics. In a similar manner we can diagonalize the second term making use of monopole vector spherical harmonics.  The angular momentum operator components, $L_x$, $L_y$, $L_z$ can be defined 
 as in section~\ref{QM} and the vector nature of the gauge field is taken into account by using a spin-one basis for the coordinate functions. Effectively, we are thus searching for  eigenfunctions corresponding to the total angular momentum $J=L+S$ with the
 spin eigenvalue equal to one. We have already seen that in the monopole background with $B=1$ the possible values of the
 angular momentum is $\ell=\frac{1}{2}, \frac{3}{2},\ldots$ which means that we have the following possibilities for the total 
 angular momentum, $J$
 \begin{eqnarray}
 \mbox{For } \ell \geq \frac{3}{2}: &&\hspace{0.0cm} J=\ell-1,\ell,\ell+1, \\
  \mbox{For } \ell =\frac{1}{2}:&&\hspace{0.0cm} J=\ell,\ell+1.
 \end{eqnarray}
 Combining the monopole spherical harmonics with $S=1$ representation matrices one can construct monopole vector spherical harmonics~\cite{Olsen:1990jm} which fulfill
 \begin{eqnarray}
 (L+S)^2 \, \Yvec &= & J(J+1) \Yvec, \\
 (L_z+S_z) \, \Yvec&=& M\, \Yvec, \\
 L^2 \Yvec &=& \ell(\ell+1) \Yvec, \label{L2vec} \\
 S^2 \Yvec &=& 2\, \Yvec.
 \end{eqnarray}
 These vector spherical harmonics have the following additional properties which will prove convenient in the following
 \begin{eqnarray}
 {\bf \hat{r}}\cdot \Yvec &=& {\cal C}_{J\ell}^{(q)} \, Y_{J M}^{(q)}(\theta,\phi), \label{rdot} \\
 {\bf \hat{r}}\times \Yvec &=&i\sum_L {\cal A}_{J\ell L}^{(q)} \,  {{\bf Y}_{JL M}^{(q)}(\theta,\phi)}, \label{rcross}
 \end{eqnarray}
 where the constants ${\cal A}_{J\ell L}^{(q)}$ and ${\cal C}_{J\ell}^{(q)}$ are known in closed form. 
 We have listed these in appendix~\ref{Clebsch}.
 In addition, one has
 \begin{equation}
 {\bf r} \, Y_{JM}^{(q)} = \sum_\ell C_{J\ell}^{(q)} \,\Yvec.
 \end{equation}
  We thus expand the scalar field in the basis of monopole spherical harmonics and the vector field in the basis of monopole
 vector spherical harmonics in the following way (leaving out the subscripts on the fields)
 \begin{eqnarray}
 \varphi (r,\theta,\phi)&=&\sum_{J=\frac{1}{2},\frac{3}{2},\ldots} \, \sum_{M=-J}^J r^{-1} \,K^J(r) \,Y_{JM}^{(q)}(\theta,\phi), \\
 \vec{A}(r,\theta,\phi)&=&\sum_{J=\frac{1}{2},\frac{3}{2},\ldots} \, \sum_{M=-J}^J \vec{A}_{JM}(r,\theta,\phi),
  \end{eqnarray}
where
\begin{eqnarray}
\lefteqn{
r\vec{A}_{JM}(r,\theta,\phi)=} \\
&&F_-^J(r) {{\bf Y}_{JJ-1 M}^{(q)}(\theta,\phi)}+F_0^J(r){{\bf Y}_{JJ M}^{(q)}(\theta,\phi)}+F_+^J(r){{\bf Y}_{JJ+1 M}^{(q)}(\theta,\phi)},
 \nonumber
\end{eqnarray}
with the understanding that $F_-^{\frac{1}{2}}(r)\equiv0$. Working in this basis the spectral
 problem 
that we need to solve reads
\begin{equation}
\hat{H}\begin{pmatrix} \varphi \\ \vec{A}
\end{pmatrix} \,=\, \frac{1}{r^2}
\begin{pmatrix}
r^2 p_r^2+{\bf L}^2 & -i B \,{\bf \hat{r}^T} \\
iB {\bf \hat r}&  r^2 p_r^2+{\bf L}^2 -iB \, {\bf \hat{r} \times} \end{pmatrix} 
\begin{pmatrix} \varphi \\ \vec{A}
\end{pmatrix}
=E\begin{pmatrix} \varphi \\ \vec{A} \label{mixingproblem}
\end{pmatrix}.
\end{equation}
We will follow the strategy of~\cite{Olsen:1990jm} where the part of the problem involving only the gauge field 
was solved.
Inserting the expansions of $\varphi$ and $\vec{A}$ and making use of the relations~(\ref{L2vec}), (\ref{rdot}), and (\ref{rcross})
one finds for each value of $J$ a set of four coupled differential equations for the radial functions $F_-^J(r)$, 
$F_0^J(r)$, $F_+^J(r)$ and
$K^J(r)$. The first of these equations reads
\begin{eqnarray}
\lefteqn{
\left(z^2 \frac{d^2}{dz^2} +z^2-J(J+1)\right )K(z) } \\
&&+iB \left({\cal C}_{JJ-1}^{(q)}\,F_-^J(z)+{\cal C}_{JJ}^{(q)}\,F^J_0(z)+ {\cal C}_{JJ+1}^{(q)}\,F^J_+(z)
\right)=0, \nonumber
\end{eqnarray}
where we have defined
\begin{equation}
z=\sqrt{E}\,r.
\end{equation}
The other three equations are analogous.  We
now notice that the function $\sqrt{z}J_{\nu}(z)$ fulfils the equation
\begin{equation}
\left(z^2 \frac{d^2}{dz^2}+z^2-\left(\nu^2-\frac{1}{4}\right)\right)\sqrt{z}J_{\nu}(z)=0.
\end{equation}
It thus makes sense to search for a solution to the set of four coupled differential equations where all the four
unknown functions are  proportional to $\sqrt{z} J_{\nu}(z)$ for some $\nu$.
Writing
\begin{equation}
K^J(z)= K^J \sqrt{z} J_{\nu}(z), \nonumber
\end{equation}
\begin{equation}
F_-^J(z)= F_-^J \sqrt{z} J_{\nu}(z),\hspace{0,5cm} F_0^J(z)= F_0^J \sqrt{z} J_{\nu}(z),\hspace{0.5cm}
F_+^J(z)= F_+^J \sqrt{z} J_{\nu}(z), \nonumber
\end{equation}
where $K^J$, $F_-^J$, $F_0^J$ and $F_+^J$ are constants we get a set of four linear equations for these four unknown
constants, which we can write as
\begin{equation}
\begin{bmatrix}
a &  i \,{\cal C}_{J-1} &  i \,{\cal C}_{J} & i \,{\cal C}_{J+1} \\
- i \,{\cal C}_{J-1}& a+2J -\,{\cal A}_{J-1J-1}& -\,{\cal A}_{ J J-1} & 0 \\
  -i {\cal C}_{J} & -\,{\cal A} _{J-1J} & a-\,{\cal A}_{JJ} &  -\,{\cal A}_{J+1J} \\
   -i C_{ J+1} &0 &  -\,{\cal A}_{J J+1} & a-2(J+1) - {\cal A}_{ J+1J+1}
\end{bmatrix} \nonumber
\begin{bmatrix}
K\\
F_- \\
F_0 \\
F_+ 
\end{bmatrix}= \begin{bmatrix}
0 \\
0 \\
0 \\
0 
\end{bmatrix},
\end{equation}
where for readability we have suppressed the superscript $q$ as well as the first indices on the ${\cal C}$'s and ${\cal A}$'s which obviously all take the
value $J$, and likewise for the superscripts on the constants. Furthermore, we have introduced
\begin{equation}
a=\nu^2-\left(J+\frac{1}{2}\right)^2,
\end{equation}
and we have set $B=1$.
The possible values of $\nu$ for a given $J$ are thus given by the values of $a$ for which the determinant of the above
matrix vanishes (i.e.\ by minus the eigenvalues of the matrix), which are 
\begin{equation}
a=\left\{0,0,-2J,2(J+1)\right\}, \hspace{0.5cm} J\geq 3/2,
\end{equation}
corresponding to 
\begin{equation}
\nu=\left\{J-\frac{1}{2},J+\frac{1}{2},J+\frac{1}{2},J+\frac{3}{2}\right\}\equiv\{\nu^J_-,\nu^J_0,\nu^J_0,\nu^J_+\}, \hspace{0.5cm} J\geq 3/2. \label{nus}
\end{equation}
It is interesting to note the difference between this result and the result of~\cite{Olsen:1990jm}. Of the three values of $\nu$, which
one gets in the situation where one considers  only the gauge field, one value equals $\nu_0^J$ but the other two are irrational.
The coupling to the scalar field, originating from the supersymmetry of ${\cal N}=4$ SYM, clearly renders the problem 
analytically more tractable, manifesting the presence of integrability.
We can choose a set of orthonormal eigenvectors of the above matrix as follows
\begin{eqnarray}
V_-^J&=& \left(\frac{-i}{2\sqrt{J(2J+1)}},\frac{\sqrt{(2J+1)(2J-1)}}{2J}, \frac{1}{2J}\sqrt{\frac{J+1}{2J+1}},0 \label{V-}
\right), \\V_0^J&=&\left (
-\frac{i}{2}\sqrt{\frac{2J-1}{J}}, -\frac{1}{2J},\frac{\sqrt{(2J-1)(J+1)}}{2J},0  \label{V0}
\right), \nonumber\\
\widehat{V}_0^J&=& \left(\frac{i}{2}\sqrt{\frac{2J+3}{J+1}},0, \frac{\sqrt{J(2J+3)}}{2J+2},\frac{1}{2J+2}  \label{V0hat}
\right), \nonumber \\
V_+^J&=& \left(\frac{-i}{2\sqrt{(J+1)(2J+1)}},0,
\frac{-J}{2(1+J)\sqrt{J(1+2J)}},\frac{\sqrt{(2J+1)(2J+3)}}{2(1+J)} \label{V+} \nonumber
\right).
\end{eqnarray}
For the special case of $J=1/2$, where $F_-^{\frac{1}{2}}(r)\equiv 0$, we find
\begin{equation}
\nu=\{0,1,2\}\equiv\{\nu^{\frac{1}{2}}_-,\nu^{\frac{1}{2}}_0,\nu^{\frac{1}{2}}_+\}, \hspace{0.5cm} J=1/2,
\end{equation}
with the corresponding eigenvectors
\begin{eqnarray}
V_-^{\frac{1}{2}}&=&\left(-\frac{i}{2},\sqrt{\frac{3}{4}},0
\right),\hspace{0.4cm}V_0^{\frac{1}{2}}=\left(i\sqrt{\frac{2}{3}},\sqrt{\frac{2}{3}},\frac{1}{3} \nonumber
\right),\hspace{0.4cm}
  \\
V_+^{\frac{1}{2}}&=&\left(\frac{-i}{2\sqrt{3}},-\frac{1}{6},\frac{2\sqrt{2}}{3}, \label{Vspecial2}
\right),
\end{eqnarray}
consistently with the above. These eigenvectors provide us with a basis for the space of solutions of the quantum mechanical
problem~(\ref{mixingproblem}) and we can determine the propagators of the fields with flavour mixing.  In the following we
will explicitly determine only the propagators for the scalars, but it is not difficult to reconstruct the vector propagator as well. The fermion propagators are naturally expanded in the spinor monopole harmonics \cite{Kazama:1976fm}.

 \subsection{The propagators for the scalars \label{Propagators}}
 
 The equation defining the propagators of the simple scalars, cf.\ eqn.~\eqref{Lsimple} reads (assuming Euclidean signature)
\begin{eqnarray}
\lefteqn{\left(-\frac{\partial^2}{\partial t^2}-\frac{\partial^2}{\partial r^2}-\frac{2}{r} \frac{\partial}{\partial r} +\frac{1}{r^2} L^2(\theta,\phi)\right)
G({\bf{x}},{\bf{x'}})=} \nonumber\\
&&\hspace*{3.0cm}\frac{g_{\mbox{\tiny YM}}^2}{2} \frac{\delta(t-t')\delta(r-r')\delta(\theta-\theta')\delta(\phi-\phi')}{r^2 \sin\theta},
 \end{eqnarray}
  where $L^2(\theta,\phi)$ was defined in~\eqref{L2def} and where ${\bf{x}}=(t,r,\theta,\phi)$.
 We now make the following ansatz for $G({\bf{x}},{\bf{x'}})$
 \begin{equation}
 G({\bf{x}},{\bf{x'}})=\frac{g_{\mbox{\tiny YM}}^2}{2} \sum_{l,\bar{m}}\, (r\,r')^{-1} G_{l}(r,r',t,t')\, Y^{(q)}_{l,\bar{m}}(\theta,\phi)
 Y^{(q)*}_{l,\bar{m}}(\theta',\phi'), \label{propagator}
 \end{equation}
 and find that $G_l(r,r',t,t')$ has to fulfill the following equation
 \begin{equation}
 \left(-\frac{\partial^2}{\partial t^2}-\frac{\partial^2}{\partial r^2}+\frac{l(l+1)}{r^2}\right) G_l(r,r',t,t')=\delta(r-r')\delta(t-t'),
 \label{AdS-part}
 \end{equation}
 which we recognize as the propagator of a scalar in AdS$_2$ with mass parameter $m^2=l(l+1)$. The appearance of
 an AdS$_2$ propagator was to be expected given the symmetries of the problem.  We elaborate on this connection to AdS$_2$ in appendix~\ref{AdS2}.
 A convenient way to write the scalar propagator with mass
parameter $m^2$ for 
 AdS$ _{d+1}$ is exactly by means of its spectral decomposition~\cite{Liu:1998ty}
\begin{eqnarray}
G_\nu(r,r',\vec{x},\vec{y})&=&(r r')^{d/2} \int \frac{d^d k}{(2\pi)^d}\,  \int_0^{\infty} d \omega \frac{\omega}{\omega^2+\vec{k}^2}
e^{i \vec{k}(\vec{x}-\vec{y})} J_{\nu}(\omega r) J_\nu(\omega r')
 \nonumber  \\
&=& (r r')^{d/2} \int \frac{d^d k}{(2\pi)^d}\, e^{i\vec{k}(\vec{x}-\vec{y})}
I_{\nu}(|\vec{k}| r^<) K_{\nu}(|\vec{k}|r^>), \label{spectral}
\end{eqnarray}
where $I_\nu$ and $K_\nu$ are modified Bessel functions
and where
\begin{equation}\label{nu-def}
\nu=\sqrt{m^2+\frac{1}{4}}.
\end{equation}
For the scalar which mixes with the gauge field the relevant  mode expansion is not in terms of orbital but in terms of  total angular momentum, $J$. To get the propagator for this field we write a spectral decomposition like  (\ref{spectral}) for the propagator of the combined  field $(\varphi_1,\vec{A})$ making use of the complete basis of eigenfunctions determined earlier
and identify the desired propagator as the 11-component. This leads to an expression similar to (\ref{propagator}) 
except that the single term $G_{l}$ gets replaced with the  sum 
\begin{equation}
|(V_-^J)_1|^2G_{\nu_-^J}+  \left(|(V_0^J)_1|^2+|(\widehat{V}_0^J)_1|^2\right)
 G_{\nu_0^J}+|(V_+^J)_1|^2G_{\nu_+^J}, \nonumber
\end{equation}
where $(V_-^J)_1$ is the first component of the eigenvector $V_-^J$ etc., cf.\ eqns. (\ref{V-}-\ref{Vspecial2}), and where
it is understood that $|(\widehat{V}_0^{\frac{1}{2}})_1|=0$. 
We notice that the lowest mode which appears is $\nu_-^{1/2}$ which exactly saturates the BF-bound~\cite{Breitenlohner:1982jf}. 
 
 For the computation of  one-loop corrections to  one-point functions we will need to evaluate propagators with coinciding endpoints and we need a regularization procedure. 
For the AdS$_2$ part of the propagator we can use the method successfully employed
in~\cite{Buhl-Mortensen:2016pxs,Buhl-Mortensen:2016jqo}, i.e.\ dimensional regularization. We regularize the propagator for 
coinciding points by going to polar coordinates in $d=1-2\epsilon$ dimensions and obtain
\begin{equation}
G_{\nu}(r,r,t,t)=\frac{g_{\mbox {\tiny YM}}^2}{2} \, r\,\frac{2 \pi^{1/2-\epsilon}}{\Gamma(1/2-\epsilon)}
\int_0^{\infty}dk \, \frac{ k^{-2\epsilon}}{( 2\pi)^{1-2\epsilon}} \, I_\nu( kr) K_{\nu}(kr).
\end{equation}
Expanding to two leading orders in $\epsilon$ we get
\begin{equation}\label{regG}
G_{\nu}(r,r,t,t)=\frac{g_{\mbox {\tiny YM}}^2}{2} 
\frac{1}{4\pi} \left( \frac{1}{\epsilon}-\gamma +\log(4\pi)+2\log(r) -2\Psi(\nu+\frac{1}{2})\right). 
\end{equation}
Alternatively one can use propagators in $AdS_2$ and regularize by point-splitting. This gives the same result, see appendix~\ref{AdS2}.

\subsection{The correction to the one-point function }

In order to compute the one-loop correction to the one-point function we start from the properly normalized operator~\eqref{O_L},
insert the complex scalar $Z$ and expand around the classical field, ie.
\begin{equation}
Z=\Phi_1 +\varphi_1+i \varphi_2,
\end{equation}
with $\Phi_1$ given in eqn.~\eqref{Phi_cl} and
$\varphi_1$ and $\varphi_2$ the quantum fields. To one-loop order we only get contributions from 
configurations in the form of tadpoles where two 
quantum fields occupying neighboring sites get connected by a propagator, fig.~\ref{planar-fig}. The fields propagating in the loop are from the  $\kkb$ and $\knb$ blocks in (\ref{Decomposition}) and have color indices $\varphi _a^{1j}$, but  the contribution of the $11$ component is unaffected by the monopole and vanishes by chirality.  The non-planar diagrams~\ref{non-planar-fig} are not only large-$N$ suppressed, but identically vanish, because the fields propagating in the loop are $\varphi _a^{11}$.

In principle one could also get 
contributions from diagrams, denoted as ``lollipop'' diagrams in~\cite{Buhl-Mortensen:2016pxs,Buhl-Mortensen:2016jqo}, where a single quantum field gets connected
by a propagator to a closed loop. However, the contributions from such diagrams are expected to vanish due to supersymmetry
provided a supersymmetric regulator like dimensional regularization in combination with dimensional reduction is applied.  This was demonstrated by explicit calculation
for the supersymmetric D3-D5 probe brane set-up in~\cite{Buhl-Mortensen:2016pxs,Buhl-Mortensen:2016jqo}. Accordingly, we shall therefore ignore these diagrams.

There are $L$ tadpole configurations and $N-1$ field 
components with a non-trivial  coupling to $\Phi_1$. This implies
\begin{eqnarray}
\lefteqn{\langle{\cal O}_L ({\bf{x}})\rangle_T=\frac{1}{\sqrt{L}} \left(\frac{4 \pi^2}{\lambda}\right)^{L/2} \times }\\
&&\left\{\frac{1}{(2 r)^L}+\frac{(N-1) L}{(2 r)^{L-2}}  \left[\langle \varphi_1(x)\varphi_1(x) \rangle -\langle \varphi_2(x)\varphi_2(x) \rangle \right]
\right\}.  \nonumber 
\end{eqnarray}
Making use of the identity
\begin{equation}
\sum_M Y_{JM}^{(q)}(\theta,\phi)Y^{(q)*}_{JM}(\theta,\phi)=\frac{(2J+1)}{4\pi},
\end{equation}
we can write the propagator for the simple field at coinciding points as (cf.\ eqn.\ (\ref{propagator}))
\begin{equation}
\langle \varphi_2(x)\varphi_2(x) \rangle= \frac{g_{\mbox {\tiny YM}}^2}{2} \frac{1}{r^2} \sum_{J=\frac{1}{2},\frac{3}{2},\ldots}\frac{2J+1}{4\pi}\,\,G_{\nu_0^J}(x,x). \label{propsum}
\end{equation}
For the field $\varphi_1$ which mixes with the gauge field  we get a similar expression
except that the single term $G_{\nu_0^J}(x,x)$ gets replaced with the  sum 
\begin{equation}
|(V_-^J)_1|^2G_{\nu_-^J}(x,x)+  \left(|(V_0^J)_1|^2+|(\widehat{V}_0^J)_1|^2\right)
 G_{\nu_0^J}(x,x)+|(V_+^J)_1|^2G_{\nu_+^J}(x,x). \nonumber
\end{equation}
Noting that the following relation is fulfilled
\begin{equation}
|(V_-^J)_1|^2+|(V_0^J)_1|^2+|(\widehat{V}_0^J)_1|^2+|(V_+^J)_1|^2=1,
\end{equation}
we see that all terms containing a pole in $\epsilon$ cancel out and we are left with the following expression for the
one-loop correction to the one-point function
\begin{eqnarray}
\langle{\cal O}_L ({\bf{x}})\rangle_T^{1-loop}&=&
\frac{1}{\sqrt{L}} \left(\frac{4 \pi^2}{\lambda}\right)^{L/2} 
\left\{ \frac{1}{(2r)^L}\frac{g_{\footnotesize YM} ^2(N-1)L}{4 \pi^2} 
\,\times \right. \nonumber \\
&&
\left. \sum_J (2J+1) \left[
 \left(1-|(V_0^J)_1|^2-|(\widehat{V}_0^J)_1|^2
 \right)\Psi\left( \nu_0^J+\frac{1}{2}\right)  \right. \right. \nonumber\\
&&\left.  \left.-|(V_-^J)_1|^2\,
 \Psi\left(\nu_- +\frac{1}{2}\right)-
 |(V_+^J)_1|^2\,
 \Psi\left(\nu_++\frac{1}{2}\right)
\right]
\right\}.
\end{eqnarray}
Furthermore, upon using the identity
\begin{equation}
\Psi(z+1)=\Psi(z)+\frac{1}{z},
\end{equation}
as well as the exact expressions for the relevant components of the eigenvectors,
the sum telescopes and we obtain:
\begin{eqnarray}
\langle{\cal O}_L ({\bf{x}})\rangle_T^{1-loop}&=&
\frac{1}{\sqrt{L}} \left(\frac{4 \pi^2}{\lambda}\right)^{L/2} \frac{1}{(2r)^L}
\left\{ \frac{g_{\footnotesize YM} ^2(N-1) L}{4 \pi^2} 
\,\sum_{J=\frac{1}{2},\frac{3}{2},\ldots}  \, \frac{1}{4}\left(\frac{1}{J^2}-\frac{1}{(J+1)^2}\right) \right\} \nonumber \\
&=&\frac{1}{\sqrt{L}} \left(\frac{4 \pi^2}{\lambda}\right)^{L/2} \frac{1}{(2r)^L}
\left( \frac{g_{\footnotesize YM} ^2(N-1) L}{4 \pi^2} \right),
\end{eqnarray}
exactly as predicted by~(\ref{prediction}).

\section{Unprotected operators and integrability \label{integrability}}

As we discussed in the introduction the straight 't~Hooft line is expected to preserve integrability of $\mathcal{N}=4$ SYM. The evidence is based on classical string theory where the 't~Hooft line is described by a D1-brane in the bulk of $AdS_5\times S^5$ that features integrability-preserving boundary conditions \cite{Dekel:2011ja}. Here we will study integrability from the weak-coupling perspective by probing the 't~Hooft line with local operators. Single-trace operators can be viewed as states in an integrable quantum spin chain, while a D-brane (the 't~Hooft line in this case) maps to a boundary state whose overlaps with the Bethe eigenstates determine the one-point correlation functions \cite{deLeeuw:2015hxa,Buhl-Mortensen:2015gfd}. The boundary state may or may not be integrable. If it is, the one-point functions admit a compact determinant representation \cite{deLeeuw:2015hxa,Buhl-Mortensen:2015gfd} that potentially can be extended to all operators and all loop orders by a version of integrability bootstrap \cite{Komatsu:2020sup,Gombor:2020kgu,Gombor:2020auk}. Our goal is to establish perturbative integrability of 't~Hooft loops. We will
limit ourselves to the leading perturbative order and to specific sectors where the spin chain takes a particularly simple form. 

\subsection{$SO(6)$ sector}

The $SO(6)$ sector consists of generic scalar operators
\begin{equation}\label{genscalop}
 \mathcal{O}=\Psi ^{I_1\ldots I_L}\mathop{\mathrm{tr}}\Phi _{I_1}\ldots 
 \Phi _{I_L},
\end{equation}
which at one loop only mix among themselves. The mixing matrix acting on the wavefunction $\Psi ^{I_1\ldots I_L}$ coincides with the Hamiltonian of an integrable $SO(6)$ spin chain \cite{Minahan:2002ve}:
\begin{equation}
 \Gamma =\frac{\lambda }{16\pi ^2}\sum_{\ell=1}^{L}\left(
 2-2P_{\ell \ell+1}+K_{\ell \ell+1}
 \right),
\end{equation}
where $P$ and $K$ are permutation and trace operators acting on the neighboring sites: $P^{IK}_{JL}=\delta ^I_L\delta ^K_J$, $K^{IK}_{JL}=\delta ^{IK}\delta _{JL}$.

The model is solvable by Bethe ansatz with the standard Lie-algebraic Bethe equations:
\begin{equation}
 \left(\frac{u_{aj}-\frac{iq_a}{2}}{u_{aj}+\frac{iq_a}{2}}\right)^L
 \prod_{bk}^{}\frac{u_{aj}-u_{bk}+\frac{iM_{ab}}{2}}{u_{aj}-u_{bk}-\frac{iM_{ab}}{2}}
 =-1,
\end{equation}
where $M_{ab}$ is the Cartan matrix of $SO(6)$ and $q_a$ are the Dynkin labels of its vector representation:
\begin{equation}
 M=\begin{bmatrix}
  2 & -1  & 0  \\ 
  -1 & 2 & -1 \\ 
  0 & -1 & 2 \\ 
 \end{bmatrix},\qquad q=\begin{bmatrix}
  0 \\ 
  1 \\ 
  0 \\ 
 \end{bmatrix}.
\end{equation}
The one-loop anomalous dimension is given by
\begin{equation}
 \Delta =L+\frac{\lambda }{8\pi ^2}\sum_{aj}^{}\frac{q_{aj}^2}{u_{aj}^2+\frac{q_{aj}^2}{4}}\,,
\end{equation}
for a state with $K_a$ Bethe roots and quantum numbers (the Dynkin labels of the $SO(6)$ representation): 
\begin{equation}
 \mu _a=Lq_a-\sum_{b}^{}M_{ab}K_b.
\end{equation}

Computing the correlator of (\ref{genscalop}) with the 't~Hooft loop at the tree level is straightforward, each field is replaced by its classical expectation value (\ref{Phi_cl}).  The spin-chain wavefunction gets projected on a fixed external state as a result:
\begin{equation}\label{Bst}
 \Bst_{I_1\ldots I_L}=n_{I_1}\ldots n_{I_L}.
\end{equation}
The one-point function is the overlap with this state:
\begin{equation}
 \left\langle \mathcal{O}(x)\right\rangle_T=\left(\frac{2\pi ^2}{\lambda r^2}\right)^{\frac{L}{2}}
 L^{-\frac{1}{2}}\,\frac{\left\langle \Bst\right.\!\left| \Psi \right\rangle}{\left\langle \Psi \right.\!\left|\Psi  \right\rangle^{\frac{1}{2}}}\,.
\end{equation}
An extra $\lambda $-dependent prefactor arises because of the difference between the spin-chain norm and (\ref{field-Norm}). Propagators in the two-point function add a factor of  $\lambda /8\pi ^2$ each and an overall cyclicity gives rise to a factor of $L$.

The boundary state (\ref{Bst}) is known to be integrable  \cite{deLeeuw:2019ebw}.
Indeed, the boundary state picks up a single component of the wavefunction, say $\Psi ^{1\ldots 1}$ for $n=(1,\mathbf{0})$. Such a projection can be shown to commute with all odd charges of the integrable hierarchy   \cite{deLeeuw:2019ebw}, which is how integrability is defined for boundary states
  \cite{Ghoshal:1993tm,Pozsgay:2018dzs}.

  According to that definition non-zero overlaps are allowed only for parity-invariant configurations of Bethe roots:
\begin{equation}
 \left\{u_{aj}\right\}=\left\{-u_{aj}\right\}.
\end{equation}
Allowed configurations may either contain pairs of roots $(u_{aj},-u_{aj})$ or in addition include a solitary root at zero.

For states satisfying the selection rule the overlap admits an elegant determinant representation \cite{deLeeuw:2019ebw}:
\begin{equation}\label{detO(6)}
  \left\langle \mathcal{O}(x)\right\rangle_T=\left(\frac{\pi}{\sqrt{\lambda }\,r}\right)^L
  \sqrt{
  \frac{1}{L}\,\,\frac{\prod\limits_{j}^{}u_{2j}^2\left(u_{2j}^2+\frac{1}{4}\right)}{\prod\limits_{j}^{}u_{1j}^2\left(u_{1j}^2+\frac{1}{4}\right)\prod\limits_{j}^{}u_{3j}^2\left(u_{3j}^2+\frac{1}{4}\right)}\,\,
  \frac{\det G^+}{\det G^-}
  }\,,
\end{equation}
where the products are over half of the roots (picking one root in each pair), and the Gaudin factors are $K/2\times K/2$ determinants:
\begin{eqnarray}
 G^\pm_{aj,bk}&=&\left(\frac{Lq_a}{u_{aj}^2+\frac{1}{4}}-\sum_{cl}^{}K^+_{aj,cl}\right)\delta _{ab}\delta _{jk}+K^\pm_{aj,bk},
\nonumber \\
 K_{aj,bk}^\pm&=&\frac{M_{ab}}{(u_{aj}-u_{bk})^2+\frac{M_{ab}^2}{4}}\pm\frac{M_{ab}}{(u_{aj}+u_{bk})^2+\frac{M_{ab}^2}{4}}\,,
\end{eqnarray}
with indices labelling positive roots again. For configurations containing zero roots or singular roots at $\pm i/2$ the determinant formula needs to be slightly modified. The general expression that accommodates these special cases can be found in \cite{Kristjansen:2021xno}.

\subsection{$SL(2)$ sector}

The $SL(2)$ sector consists of operators
\begin{equation}
 \mathcal{O}=\sum_{\left\{n_l\right\}}^{}\frac{1}{n_1!\ldots n_L!}\,\Psi _{n_1\ldots n_L}
 \mathop{\mathrm{tr}}D^{n_1}Z\ldots D^{n_L}Z,\qquad D\equiv D_0+D_3.
\end{equation}
The states at each site are conventionally labelled as
\begin{equation}
 \left|n\right\rangle ~\longleftrightarrow~\frac{1}{n!}\,D^{n}Z.
\end{equation}
They form an infinite-dimensional unitary representation of $\mathfrak{sl}(2)$:
\begin{eqnarray}
 D\left|n\right\rangle&=&(n+1)\left|n+1\right\rangle,
\nonumber \\
 K\left|n\right\rangle&=&n\left|n-1\right\rangle,
\nonumber \\
S\left|n\right\rangle&=&\left(n+\frac{1}{2}\right)\left|n\right\rangle,
\end{eqnarray}
where $K=D^\dagger $ generates special conformal transformations and $S$ is the spin.

Operator mixing in the $\mathfrak{sl}(2)$ sector is described by the spin-chain Hamiltonian
\begin{equation}
 \Gamma =\frac{\lambda }{8\pi ^2}\sum_{\ell=1}^{L}h_{\ell,\ell+1},
\end{equation}
with  \cite{Beisert:2003jj}
\begin{equation}
 h\left|n,m\right\rangle=\sum_{k}^{}\left[
 \left(H(n)+H(m)\right)\delta _{kn}-\frac{1-\delta _{kn}}{|k-n|}
 \right]\left|k,n+m-k\right\rangle,
\end{equation}
where $H(n)$ is the harmonic number:
\begin{equation}
 H(n)=\sum_{s=1}^{n}\frac{1}{s}\,.
\end{equation}
The spectral equations are
\begin{equation}
 \left(\frac{u_j-\frac{i}{2}}{u_j+\frac{i}{2}}\right)^L\prod_{k}{\frac{u_j-u_k-i}{u_j-u_k+i}}=-1,\qquad E=\frac{\lambda }{8\pi ^2}\sum_{j}^{}\frac{1}{u_j^2+\frac{1}{4}}\,.
\end{equation}

To find the one-point function we just need to substitute the classical expectation value (\ref{Phi_cl}) for $Z$. The derivative $D$ acts on $Z$ as $\partial _3$, and we can use the identity
\begin{equation}
 \partial _3^n\,\frac{1}{r}=\frac{(-1)^nn!}{r^{n+1}}\,P_n(\cos\theta )
\end{equation}
to express the result of differentiation on each site. Here $\theta $ is the angle between the radius vector and the $x_3$ direction, and $P_n$ are Legendre polynomials. The one-point function can then be written in the overlap form:
\begin{equation}\label{<O>-explicit-sl2}
 \left\langle \mathcal{O}(x)\right\rangle_T
 =\left(\frac{2\pi}{\sqrt{\lambda} }\right)^{L}\frac{L^{-\frac{1}{2}}}{r^{L+S}}\,\,\frac{\left\langle {\rm Bsl}\right.\!\left| \Psi \right\rangle}{\left\langle \Psi \right.\!\left|\Psi  \right\rangle^\frac{1}{2}}\,,
\end{equation}
where $S=n_1+\ldots +n_L$ and
\begin{equation}\label{Bsl-def}
 \left\langle {\rm Bsl}\right|=\left\langle B \right|\otimes\ldots \otimes \left\langle B\right|,
\end{equation}
with
\begin{equation}\label{B-def}
 \left\langle B\right|=\sum_{n=0}^{\infty }(-1)^n
P_n(\cos\theta )\,\left\langle n\right|
=
\left\langle 0\right|
\sum_{n=0}^{\infty }\frac{(-K)^n}{n!}\,
P_n(\cos\theta ).
\end{equation}
We posit that this is an integrable boundary state.

In fact, at $\theta =0$ this state is just a vacuum descendant. For $\theta =0$ we have $P_n(1)=1$ and
\begin{equation}
 \left.\vphantom{\frac{1}{2}}\left\langle B\right|\right|_{\theta =0}
 =\left\langle 0\right|\,{\rm e}\,^{-K}.
\end{equation}
The additivity of the exponential then implies
\begin{equation}
\left.\vphantom{\frac{1}{2}}\left\langle {\rm Bsl}\right|\right|_{\theta =0}
=\left\langle 0\right|\,{\rm e}\,^{-{K_{\rm tot}}},
\end{equation}
where
\begin{equation}
 K_{\rm tot}=\sum_{l=1}^{L}K_l.
\end{equation}
And $K_{\rm tot}=D_{\rm tot}^\dagger $.

The special conformal generator annihilates any primary state:
\begin{equation}
 K_{\rm tot}\left|\Psi_{\rm primary} \right\rangle=0.
\end{equation}
Hence, the one-point function of any conformal primary vanishes on the positive $x_3$ semi-axis:
\begin{equation}
 \left.\vphantom{\frac{1}{2}}\left\langle O_{\rm primary}(x)\right\rangle_T\right|_{\theta =0}=0,
\end{equation}
the only exception being the spin-chain vacuum $\mathop{\mathrm{tr}}Z^L$.

The one-point function in general position is not protected by symmetries and in general does not vanish. We find that it is described by the following determinant formula:
\begin{equation}\label{sl2-det}
 \left\langle \mathcal{O}(x)\right\rangle_T
 =\left(\frac{2\pi }{\sqrt{\lambda }\,r}\right)^L\left(\frac{\sin\theta }{r}\right)^S\sqrt{\frac{1}{L}\,\,\frac{Q(0)}{Q\left(\frac{i}{2}\right)}\,\,\frac{\det G^+}{\det G^-}}\,,
\end{equation}
which applies to any state with paired roots\footnote{A root at zero is also allowed and can be included by appropriately modifying the Gaudin factors \cite{Brockmann:2014b}. For unpaired states the one-point function must vanish.}: $\left\{u_j,-u_j\right\}_{j=1\ldots \frac{S}{2}}$. The Q-functions are defined as
\begin{equation}
 Q(x)=\prod_{j=1}^{\frac{S}{2}}\left(x^2-u_j^2\right),
\end{equation}
and the Gaudin factors are given by
\begin{eqnarray}
 G^\pm_{jk}&=&\delta _{jk}\left(\frac{L}{u_j^2+\frac{1}{4}}-\sum_{s}^{}K^+_{js}\right)+K^\pm_{jk},
\nonumber \\
K^\pm_{jk}&=&-\frac{2}{(u_j-u_k)^2+1}\mp \frac{2}{(u_j+u_k)^2+1}\,.
\end{eqnarray}
We  were lead to this formula by taking inspiration from general results on integrable overlaps and thoroughly checked the formula on a few dozens of eigenstates. In addition, we checked that the overlap vanishes for unpaired configurations of Bethe roots.
Explicit formulas of this type for the $\mathfrak{sl}(2)$ spin chain, but for different types of boundary states, have been presented in~\cite{Jiang:2019xdz,Jiang:2020sdw,Kristjansen:2020mhn}. 

The most powerful approach to first-principles derivation of overlap formulas is based on the algebraic Bethe ansatz \cite{Gombor:2019bun,Gombor:2020kgu,Gombor:2021uxz,Gombor:2021hmj} and is  applicable to spin chains with an infinite-dimensional Hilbert space, such as the $\mathfrak{sl}(2)$ spin chain at hand. We were informed by Tamas Gombor that  the boundary state (\ref{Bsl-def}), (\ref{B-def}) falls into the general category of integrable boundary states considered in \cite{Gombor:2021uxz} and that the overlap formula (\ref{sl2-det}) can be derived by the algebraic methods presented there\footnote{T.~Gombor, private communication.}.

For illustration of the overlap representation of the one-point function, consider the two-magnon states \cite{Staudacher:2004tk}:
\begin{equation}\label{2-magnon-sl2}
 \left|\Psi \right\rangle=\left\{
 \sum_{\ell<\ell'}^{}\left[\,{\rm e}\,^{ip(\ell-\ell')}+\,{\rm e}\,^{ip(\ell'-\ell)+i\delta }\right]D(\ell)D(\ell')
 +\cos\frac{\delta }{2}\,\,{\rm e}\,^{\frac{i\delta }{2}}
 \sum_{\ell}^{}D(\ell)^2
 \right\}\left|0\right\rangle.
\end{equation}
These are Bethe eigenstates with two Bethe roots $\left\{u,-u\right\}$, where
\begin{equation}\label{updelta}
 \,{\rm e}\,^{ip}=\frac{u+\frac{i}{2}}{u-\frac{i}{2}}=\,{\rm e}\,^{i\delta }.
\end{equation}

The following sums appear in the overlap and the norm:
\begin{eqnarray}\label{sums-2-magnons}
 \sum_{\ell<\ell'}^{}\left[
 \,{\rm e}\,^{ip(\ell-\ell')}+\,{\rm e}\,^{ip(\ell'-\ell)+i\delta }
 \right]&=&
 -L\,{\rm e}\,^{\frac{i\delta }{2}}\left(\cos\frac{\delta }{2}+\cot\frac{p}{2}\,\sin\frac{\delta }{2}\right),
\nonumber \\
 \sum_{\ell<\ell'}^{}\left|
 \,{\rm e}\,^{ip(\ell-\ell')}+\,{\rm e}\,^{ip(\ell'-\ell)+i\delta} 
 \right|^2&=&L\left(L-1-\cos\delta -\cot p\,\sin\delta \right),
\end{eqnarray}
where the Bethe equation $\,{\rm e}\,^{ipL+i\delta }=1$ has been used to simplify the expressions. For the norm and the overlap we get:
\begin{eqnarray}
 \left\langle \Psi \right.\!\left| \Psi \right\rangle&=&L(L+1),
\nonumber \\
 \left\langle {\rm Bsl}\right.\!\left|\Psi  \right\rangle&=&-\frac{L}{r^2}\,
 \cos\frac{p }{2}\,\,{\rm e}\,^{\frac{i\delta }{2}}\sin^2\theta .
\end{eqnarray}
Here we also used (\ref{updelta}).
Dropping a phase we then get for the one-point function in (\ref{<O>-explicit-sl2}):
\begin{equation}
 \left\langle \mathcal{O}(x)\right\rangle_T
 =\left(\frac{2\pi }{\sqrt{\lambda }\,r}\right)^L\left(\frac{\sin\theta }{r}\right)^2\frac{u}{\sqrt{(L+1)\left(u^2+\frac{1}{4}\right)}}\,.
\end{equation}

For the state with two roots the Gaudin factors are just numbers:
\begin{equation}
 G^+=\frac{L}{u^2+\frac{1}{4}}\,,\qquad 
 G^-=\frac{L+1}{u^2+\frac{1}{4}}\,,
\end{equation}
and the determinant representation (\ref{sl2-det}) recovers the above result.

\subsection{Gluon subsector}

The field tensor $F_{\mu \nu }$ contains two representations of the Lorentz group. The irreducible components are the self-dual and anti-self-dual projections  which can be singled out by contracting with the 't~Hooft symbols \cite{tHooft:1976snw}:
\begin{equation}
 \eta ^\pm_{i\mu \nu }=\varepsilon _{0i\mu \nu }\pm\delta _{0\mu }\delta _{i\nu }\mp\delta _{0\nu }\delta _{i\mu }.
\end{equation}
The 't~Hooft symbols satisfy:
\begin{equation}
 \eta ^\pm_{i\mu \nu }=\pm\frac{1}{2}\varepsilon _{\mu \nu \lambda \rho }\eta ^\pm_{i\lambda \rho },
\end{equation}
and realize an embedding of $\mathfrak{su}(2)_{L/R}$ into $\mathfrak{so}(4)=\mathfrak{su}(2)_L\oplus \mathfrak{su}(2)_R$. The index $i$ in $\eta ^+_{i\mu \nu }$ is thus the vector index of $\mathfrak{su}(2)_L$ in its spin-1 representation. In a slight abuse of notation we use the same notation as for ordinary spatial indices, the latter corresponding to the diagonal subalgebra in $\mathfrak{su}(2)_L\oplus \mathfrak{su}(2)_R$. We hope this will not lead to  confusion.

The (anti-)self-dual components  of the field strength,
\begin{equation}
 F_{i}^\pm=\frac{1}{2}\,\eta ^\pm_{i\mu \nu }F^{\mu \nu },
\end{equation}
form a closed sector under dilatations, in the sense that operators
\begin{equation}\label{generic-state-FFFF}
 \mathcal{O}=\Psi ^{i_1\ldots i_L}\mathop{\mathrm{tr}}F^\pm_{i_1}\ldots F^\pm_{i_L},
\end{equation}
mix only among themselves at one loop. Operators of this form appear naturally if the $\mathfrak{psu}(2,2|4)$ spin chain is formulated in the standard Cartan basis with a single fermion node on the Dynkin diagram, also known as the "Beast" grading \cite{Beisert:2003yb}. 

The ground state (or rather the reference state) then is a gluon operator
\begin{equation}\label{ground-gluon}
 \mathcal{O}_0=\mathop{\mathrm{tr}}(F^\pm_z)^L,
\end{equation}
where 
\begin{equation}
 F^\pm_z=F^\pm_1+iF^\pm_2.
\end{equation}
This operator is not protected and acquires a one-loop anomalous dimension $3\lambda L /8\pi ^2$ \cite{Beisert:2003jj,Beisert:2003yb}. Excited states are constructed by replacing some of the $F_z$'s by
\begin{equation}
 F^\pm_x=F^\pm_3,\qquad F_{\bar{z}}=F^\pm_1-iF^\pm_2.
\end{equation}
All of them  have lower dimension, the ground state (\ref{ground-gluon}) has the largest energy possible for a given length.
For example, the length-2 singlets
\begin{equation}\label{L=2singlet}
 \mathcal{T}^\pm=\mathop{\mathrm{tr}}[F^\pm_zF^\pm_{\bar{z}}+(F^\pm_x)^2]=\mathop{\mathrm{tr}}F_i^\pm F_i^\pm,
\end{equation}
have anomalous dimension zero. These operators are in fact linear combinations of the Lagrangian density (its gluon part) and that of the topological charge:
\begin{equation}
 \mathcal{T}^\pm=\frac{1}{2}\mathop{\mathrm{tr}}F_{\mu \nu }F^{\mu \nu }\pm
 \frac{1}{2}\mathop{\mathrm{tr}}F_{\mu \nu }\widetilde{F}^{\mu \nu }.
\end{equation}
They are not renormalized because the beta-function is zero.

The one-loop mixing matrix of the gluon operators (\ref{generic-state-FFFF}) is the Hamiltonian of an $SO(3)$ spin chain  \cite{Ferretti:2004ba,Beisert:2004fv}:
\begin{equation}\label{spin-1-H}
 \Gamma =\frac{\lambda }{16\pi ^2}\sum_{l=1}^{L}\left(5+P_{l,l+1}-2K_{l,l+1}\right),
\end{equation}
where $P$ and $K$ are permutation and trace operators acting on the $SO(3)$ indices of $F^\pm_{i_l}$. The model is integrable \cite{Zamolodchikov:1980ku,Kulish:1981gi,Reshetikhin:1986vd} and its Bethe-ansatz solution \cite{A.:1982zz,Babujian:1982ib,Babujian:1983ae} is given by:
\begin{equation}
 \left(\frac{u_j-i}{u_j+i}\right)^L\prod_{k}{\frac{u_j-u_k+i}{u_j-u_k-i}}=-1,\qquad E=\frac{\lambda }{8\pi ^2}\left(3L-\sum_{j}^{}\frac{2}{u_j^2+1}\right).
\end{equation}
The singlet state (\ref{L=2singlet}), for example, corresponds to $u=\left\{1/\sqrt{3}\,,-1/\sqrt{3}\right\}$.

The (anti-)self-dual components of the field strength are linear combinations of electric and magnetic fields $F^\pm_i=B_i\mp E_i$, and for the field of the monopole we have:
\begin{equation}
 \left\langle F^\pm_i\right\rangle_T=\frac{B}{2}\,\,\frac{x_i}{r^3}\,.
\end{equation}
To unambiguously define the one-point functions we need a proper normalization condition for tensor operators in (\ref{generic-state-FFFF}). This is to some extent arbitrary, we take:
\begin{equation}
 \left\langle \mathcal{O}^+(t,\mathbf{0})\mathcal{O}^-(0,\mathbf{0})\right\rangle=\frac{1}{t^{2\Delta }}\,.
\end{equation}
Then, taking into account
\begin{equation}
 \left\langle F_i^+(t,\mathbf{0})F_j^-(0,\mathbf{0})\right\rangle
 =\frac{\lambda \delta _{ij}}{\pi ^2t^4}\,,
\end{equation}
and substituting the classical field (\ref{F_cl}) into the operator we find:
\begin{equation}\label{1pt-overlap-gluon}
 \left\langle \mathcal{O}(x)\right\rangle_T
 =\left(\frac{\pi }{2\sqrt{\lambda }\,r^3}\right)^LL^{-\frac{1}{2}}\frac{
 \left\langle {\rm Bgl}\right.\!\left|\Psi  \right\rangle
 }{\left\langle \Psi \right.\!\left|\Psi  \right\rangle^{\frac{1}{2}}}\,.
\end{equation}
The boundary state $\left\langle {\rm Bgl}\right|$ is one-site factorizable: 
\begin{equation}\label{Bgl}
 {\rm Bgl}_{i_1\ldots i_L}=x_{i_1}\ldots x_{i_L}\,.
\end{equation}
It picks one particular component of the wave function where spins on all sites point in the same direction (that of the radius vector of the operator insertion).

We expect on general grounds that this state is integrable. As such it should have non-zero overlaps only with balanced Bethe eigenstates in which all roots are paired, and should have a determinant representation similar to (\ref{detO(6)}) and (\ref{sl2-det}). We have not proven integrability in any direct way, nor have we honestly derived the overlap formula, but we have extensively checked both.  
We computed numerically overlaps with all eigenstates up to length nine and checked that they vanish for unbalanced Bethe states. 
For overlaps with paired states $u=\{u_j,-u_j\}_{j=1\ldots M/2}$ we found a concise determinant formula\footnote{The number of Bethe roots in a balanced state needs not to be even, there are exceptional states containing a triplet of singular roots at $i,-i,0$ \cite{Hao:2013rza,Hou:2023ndn}  and an even number of paired regular roots. These states are also parity-invariant and have vanishing total momentum thus allowing for non-zero overlaps. The determinant formula is ill-defined on singular roots and has to be regularized, more details can be found in \cite{Kristjansen:2020vbe}.}:
\begin{equation}
 \left\langle \mathcal{O}(x)\right\rangle_T
 =\left(\frac{\pi }{2\sqrt{2\lambda }}\right)^L
 \frac{z^{L-M}}{r^{3L-M}}
 \sqrt{
 \frac{1}{L}\,\,
 \frac{Q(0)}{Q\left(\frac{i}{2}\right)}\,\,
 \frac{\det G^+}{\det G^-}
 }\,,
\end{equation}
where $Q(x)$ is the Baxter polynomial:
\begin{equation}
 Q(x)=\prod_{j=1}^{\frac{M}{2}}\left(x^2-u_j^2\right),
\end{equation}
and $G^\pm$ are the Gaudin factors:
\begin{eqnarray}\label{Gaudin+-}
 G^\pm_{ij}&=&\left(\frac{2L}{u_j^2+1}-\sum_{s}^{}K^+_{js}\right)\delta _{jk}+K^\pm_{jk},
 \\
K^\pm_{jk}&=&\frac{2}{(u_j-u_k)^2+1}\pm\frac{2}{(u_j+u_k)^2+1}\,.
\end{eqnarray}

The space-time dependence of the one-point function is very simple, simpler than required by explicit symmetries alone. Those require the spin-$S$ overlap to be of the form $z^S\mathcal{P}$, where $\mathcal{P}$ is a homogeneous polynomial  in $x\equiv x_3$ and $|z|$. For the highest-weight states the polynomial prefactor neatly combines into $(x^2+|z|^2)^{(L-S)/2}=r^M$. This simplification, we believe, is also a consequence of integrability.

To exemplify the use of the overlap formula, we consider the two-magnon states of arbitrary length with Bethe roots at $u$ and $-u$ constrained by the Bethe equations. The chiral components of the field strength form the canonical basis of the $\mathfrak{su}(2)$ spin-1 representation:
\begin{equation}
 F^\pm_z\longleftrightarrow\sqrt{2}\left|1\right\rangle,
 \qquad 
 F^\pm_x\longleftrightarrow\left|0\right\rangle,
 \qquad 
  F^\pm_{\bar{z}}\longleftrightarrow\sqrt{2}\left|-1\right\rangle.
\end{equation}
The spin lowering operator acts as
\begin{equation}
 L_-F^\pm_z=2F_x^\pm,
 \qquad 
 L_-F_x^\pm=-F_{\bar{z}}^\pm.
\end{equation}
The two-magnon eigenstate of the Hamiltonian (\ref{spin-1-H}) can be generated by applying the lowering operators twice:
\begin{equation}
 \left|\Psi \right\rangle=\frac{1}{2}\left\{
 \sum_{\ell<\ell'}^{}\left[
 \,{\rm e}\,^{ip(\ell-\ell')}
 +\,{\rm e}\,^{ip(\ell'-\ell)+i\delta }
 \right]L_-(\ell)L_-(\ell')
 +\cos\frac{\delta }{2}\,\,{\rm e}\,^{\frac{i\delta }{2}}
 \sum_{\ell}^{}L_-(\ell)^2
 \right\}\left|0\right\rangle,
\end{equation}
quite similarly to (\ref{2-magnon-sl2}), while the phase shift and the momentum of the magnons are now
\begin{equation}
 \,{\rm e}\,^{ip}=\frac{u+i}{u-i}\,,\qquad \,{\rm e}\,^{i\delta }=\frac{u-\frac{i}{2}}{u+\frac{i}{2}}\,.
\end{equation}

For two magnons the Gaudin factors are just numbers:
\begin{equation}
 G^+=\frac{2L}{u^2+1}\,,\qquad G^-=\frac{2L}{u^2+1}-\frac{1}{u^2+\frac{1}{4}}\,,
\end{equation}
which gives the following prediction for the one-point function:
\begin{equation}\label{O2-mag}
 \left\langle \mathcal{O}_{{\rm 2-magnon}}(x)\right\rangle_T=
 \left(\frac{\pi }{2\sqrt{2\lambda }}\right)^L\frac{z^{L-2}}{r^{3L-2}}\,\,
 \frac{u}{\sqrt{\left(u^2+\frac{1}{4}\right)\left(L-\frac{1}{2}\,\,\frac{u^2+1}{u^2+\frac{1}{4}}\right)}}\,.
\end{equation}

The the norm and the overlap can be computed with the help of (\ref{sums-2-magnons}). Taking into account the Bethe equations, we get:
\begin{eqnarray}
 \left\langle \Psi \right.\!\left| \Psi \right\rangle=2^LL\left(L-\frac{1}{2}\,\,\frac{u^2+1}{u^2+\frac{1}{4}}\right),\qquad 
 \left\langle {\rm Bgl}\right.\!\left| \Psi \right\rangle
 =Lr^2z^{L-2}\,\frac{u}{\sqrt{u^2+\frac{1}{4}}}\,.
\end{eqnarray}
 Substitution of these expressions into (\ref{1pt-overlap-gluon}) recovers (\ref{O2-mag}). 

\section{Conclusions \label{conclusion}}

Our results confirm that an infinite 't~Hooft line in the $\mathcal{N}=4$ super-Yang-Mills theory defines an integrable dCFT. Integrability opens an avenue for computing expectation values of local operators exactly without any approximations. As of now, we found exact expressions for protected operators by combining localization with S-duality. For unprotected operators our results are limited to the leading order in perturbation theory and to certain subsectors of the operator algebra, but they do reveal the integrability structure expected of the exact answer. 

The diagram technique in the monopole field, here applied to protected one-point functions, can be used for non-protected operators as well, and also to more complicated correlation functions. For example, it should be relatively straightforward to construct loop corrections to the overlap formulas we derived. 

It is not inconceivable that the full solution can be bootstrapped from integrability. Our results point to a relative simplicity of the underlying scattering theory. For comparison, scattering on the D3-D5 defect \cite{Komatsu:2020sup}  (used  to construct non-perturbative one-point functions in \cite{Komatsu:2020sup,Gombor:2020kgu,Gombor:2020auk}) involves boundary resonances. The boundary states we found have no internal structure (require no auxiliary space), and we expect the  the boundary scattering to be free of bound states. The simplicity of the protected one-point functions (\ref{1pt-Zhukowski}) points in the same direction. Analogous one-point functions for the D3-D5 system are sums of $k$ terms, each depending on $x_a$ with $a=1\ldots k$ \cite{DeLeeuw:2018cal}, plus wrapping corrections, where $k$ happens to coincide with the number of bound states in the boundary scattering~\cite{Komatsu:2020sup}. 
We only found one term that depends on $x_1$, in this sense the 't~Hooft line should be similar to the D3-D5 dCFT at $k=1$
which was studied in more detail in~\cite{Kristjansen:2020mhn}.
 
Finally, similar techniques should be applicable to monopole operators in the ABJM theory. In three dimensions monopole operators are point-like, but the one-point functions similar to those we have studied may still be non-trivial because the monopole operators and those built from the fundamental fields are not mutually local. We expect integrability of $AdS_4/CFT_3$ to be part of that story.

\subsection*{Acknowledgements}
We would like to thank T. Gombor, S.~Komatsu, G.~Linardopoulos and A.~Tseytlin for interesting discussions and J.~Gomis and R.~Izquierdo~Garcia for insightful comments.  
This work was supported by  DFF-FNU through grant number 1026-00103B (C.K.) and by VR grant 2021-04578 (K.Z.).  Nor\-dita was partially supported by Nordforsk.

\appendix

\section{One-point function at strong coupling \label{sugra}}

When
\begin{equation}
 \lambda \rightarrow \infty ,\qquad L-{\rm fixed},
\end{equation}
the correlator of the 't~Hooft loop with a chiral primary is described by classical supergravity in $AdS_5\times S^5$.

The localization prediction (\ref{CPO-OPE}) in this limit becomes
\begin{equation}\label{pred-for-sugra}
 \CC_L\simeq 2\pi \sqrt{\frac{L}{\lambda }}\,.
\end{equation}
Here we will reproduce this results from the explicit supergravity calculation.

A macroscopic object in the bulk perturbs the metric of $AdS_5\times S^5$ and  according to the standard AdS/CFT dictionary an expectation value of a local operator is the response to this perturbation as seen from the boundary. The computation of $\CC_L$ thus involves a bulk-to-boundary propagator connecting the operator insertion to the vertex operator on the world-volume of the D1-brane that represents the 't~Hooft line. We will use the method first applied to the fundamental string 
\cite{Berenstein:1998ij}  and generalized to D-branes in \cite{Bissi:2011dc}. A similar approach can be used to compute the correlator of the 't~Hooft loop with the Wilson loop \cite{Gorsky:2009pc}.

The dual of the circular 't~Hooft loop is a D1-brane of the spherical shape:
\begin{equation}\label{D1-in-AdS}
 x^\mu =\frac{R}{\cosh\tau }\,\left(\sin\sigma ,\cos\sigma ,0,0\right),
 \qquad 
 z=R\tanh\tau ,
\end{equation}
 where $(z,x^\mu )$ are the standard Poincar\'e coordinates of $AdS_5\times S^5$:
\begin{equation}
 ds^2=\frac{dx^2+dz^2}{z^2}+d\theta ^2+\sin^2\theta \,d\Omega ^2.
\end{equation}
The calculation closely follows the footsteps of \cite{Berenstein:1998ij} where the circular Wilson loop was considered.

The supergravity dual of the chiral operator (\ref{O_L}) is a scalar with $m^2=L(L-4)$ and the canonically normalized propagator\footnote{We only need the bulk-to-boundary propagator and hence send one of the points, $(w,y)$, to the boundary.}:
\begin{equation}\label{SUGRA-propagator}
\left\langle \phi _L(z,x)\phi _L(w,y)\right\rangle_{\rm SUGRA}\stackrel{w\rightarrow 0}{=}\frac{L-1}{2\pi ^2}\,\,\frac{z^Lw^L}{\left[z^2+(x-y)^2\right]^L}\,,
\end{equation}
The response function is given by the standard AdS/CFT prescription:
\begin{equation}\label{sugra-<O>}
 \left\langle \mathcal{O}_L(y)\right\rangle=\sqrt{\frac{2\pi ^2}{L-1}}\,\lim_{w\rightarrow 0}w^{-L}\left\langle \phi_L (w,y)\right\rangle_{\rm SUGRA},
\end{equation}
assuming there is a source of $\phi _L$ in the bulk.

The source in our case is the D1-brane that couples to the 10D metric through the usual DBI action:
\begin{equation}
 S_{\rm DBI}=T_{D1}\int_{}^{}d^2\sigma \,\sqrt{
 \vphantom{ \det_{ab}^{ab}}
 \det_{ab}\partial _aX^M\partial _bX^Ng_{MN}}\,.
\end{equation}
And the field $\phi _L$ resides in the metric perturbation around the ambient $AdS_5\times S^5$ background \cite{Lee:1998bxa} (see also \cite{Arutyunov:1999fb}):
\begin{eqnarray}\label{delta-g}
 \delta g_{mn}&=&2Y_L\left[
 2\nabla_m\nabla_n-L(L-1)g_{mn}
 \right]\phi _L,
\nonumber \\
 \delta g_{\alpha \beta }&=&2L(L+1)g_{\alpha \beta }Y_L\phi _L,
\nonumber \\
 a_{mnpr}&=&\varepsilon _{mnprs}(L+1)Y_L\nabla^s\phi _L,
\nonumber \\
 a_{\alpha \beta \gamma \delta }&=&-(L+1)\varepsilon _{\alpha \beta \gamma \delta \varepsilon }\nabla^\varepsilon Y_L\phi _L,
\end{eqnarray}
where $Y_L$ is the spherical function on $S^5$. For the operator at hand\footnote{The scalar fields $\Phi _i$ are in one-to-one correspondence with the Cartesian coordinates $n_i$ on $S^5$. The general chiral primary operator $C^{i_1\ldots i_L}\mathop{\mathrm{tr}}\Phi _{i_1}\ldots \Phi _{i_L}$, defined by a symmetric traceless tensor $C_A$, maps to the spherical function $Y_A=C_A^{i_1\ldots i_L}n_{i_1}\ldots n_{i_L}$. For $\mathop{\mathrm{tr}}Z^L$ the spherical function is $(n_1+in_2)^L=(\sin \theta )^L\,{\rm e}\,^{iL\varphi }$.
There are two natural ways to normalize the spherical harmonics, one by the tensor squared of $C_A$, the other by integrating the square of $Y_A$ over $S^5$.
The  $L$-dependent overall factor partly arises from this difference in normalization  \cite{Lee:1998bxa} and partly  from mixing of the $KK$ modes on $S^5$ \cite{Kim:1985ez}, it also includes the square root of the 10D gravitational constant $\kappa _{10}=(2\pi )^5/8N^2$ appearing as an overall factor in the supergravity action but absent in the canonically normalized propagator that we use to compute the amplitude.}:
\begin{equation}\label{Y_L}
 Y_L=\frac{\pi}{N\sqrt{8L(L-1)}}\,\left(\sin\theta \right)^L\,{\rm e}\,^{iL\varphi }.
\end{equation}
 
 The D1-brane does not couple directly to the four-form and resides at one point on $S^5$. Therefore only the metric fluctuations in $AdS_5$  contribute to the one-point function. Expanding the D-brane action to the linear order we find:
\begin{equation}
 \left\langle \mathcal{O}_L(y)\right\rangle_T
 =-\frac{T_{D1}}{2}\int_{}^{}d^2\sigma \,\sqrt{h}h^{ab}\partial _aX^M\partial _bX^N\delta g_{MN},
\end{equation}
where $h_{ab}$ is the induced metric on the world-volume. Taking into account (\ref{delta-g}), we get:
\begin{eqnarray}
 \left\langle \mathcal{O}_L(y)\right\rangle_T&=&-\frac{\pi ^2}{(L-1)\sqrt{\lambda L}}
 \int_{}^{}d^2\sigma \,\sqrt{h}h^{ab}\partial _aX^m\partial _bX^n
 \\
 &&\times 
 \lim_{w\rightarrow 0}w^{-L}\left\langle  \left[
 2\nabla_m\nabla_n-L(L-1)g_{mn}
 \right]\phi _L(X)\phi_L (w,y)\right\rangle_{\rm SUGRA}, \nonumber
\end{eqnarray}
where we took into account various factors appearing in (\ref{tension}), (\ref{sugra-<O>}) and (\ref{Y_L}), and set $\theta =\pi /2$, $\varphi =0$ as appropriate for the 't~Hooft line with the R-symmetry orientation $n_i=(1,\mathbf{0})$.
 
It remains to integrate the bulk-to-boundary propagator over the minimal surface  (\ref{D1-in-AdS}). The answer anticipated from the conformal symmetry is rather contrived, eq.~(\ref{genp}), but the whole dynamical information is contained in the overall constant, which can be computed by choosing any convenient $y$, for instance $y\rightarrow \infty $  \cite{Berenstein:1998ij}, as in (\ref{large-x}). Then,
\begin{equation}
 \CC_L=\frac{(L+1)\sqrt{L}}{2R^L\sqrt{\lambda }}
 \int_{}^{}d^2\sigma \,\sqrt{h}\,z^{L-2}\left[
 (\partial x)^2-(\partial z)^2
 \right].
\end{equation}
The solution (\ref{D1-in-AdS}) is in the conformal gauge, the induced metric is trivial, and we find:
\begin{equation}
 \CC_L=\frac{2\pi (L+1)\sqrt{L}}{\sqrt{\lambda }}\int_{0}^{\infty }d\tau \,\,
 \frac{\tanh^L\tau}{\cosh^2\tau }=2\pi \sqrt{\frac{L}{\lambda }}\,,
\end{equation}
in agreement with (\ref{pred-for-sugra}).

There is another regime of interest:
\begin{equation}
 \lambda \rightarrow\infty ,\qquad L\rightarrow \infty ,\qquad \frac{\lambda }{L^2}-{\rm fixed}, 
\end{equation}
the BMN limit \cite{Berenstein:2002jq}.
The localization prediction in this case is 
\begin{equation}
 \CC_L\simeq \frac{2}{\sqrt{L}}\,\sinh\frac{\pi L}{\sqrt{\lambda }}\,.
\end{equation}
The operator in the BMN limit is dual to a classical string rather than a supergravity mode. The form of the answer points to a one-loop origin, since $\sinh$ admits a product representation over $n^2+\pi ^2L^2/\lambda $, which are exactly the BMN string modes  \cite{Berenstein:2002jq}. It would be very interesting to make this argument more precise.

\section{Clebsch Gordon coefficients \label{Clebsch}}
Below we list the Clebsch Gordon coefficients entering the relations~(\ref{rdot}) and~(\ref{rcross}) for the case of 
monopole charge, $q=1/2$.
Their explicit derivation as well as their counterparts for general value of the monopole charge can be found  e.g.\ in~\cite{Olsen:1990jm}. For the ${\cal C}$'s one has
\begin{equation}
{\cal C}_{JJ-1}=\frac{1}{2} \sqrt{2-\frac{1}{J}}, \hspace{0.5cm}{\cal C}_{JJ}=-\frac{1}{2 \sqrt{J (J+1)}}, 
\hspace{0.5cm}{\cal C}_{JJ+1}=-\frac{1}{2} \sqrt{\frac{1}{J+1}+2},
\end{equation}
and for the ${\cal A}$'s one finds
\begin{align}
&{\cal A}_{J-1J-1}=\frac{1}{2 J}, \hspace{0.5cm} {\cal A}_{JJ}=-\frac{1}{2 J (J+1)}, \hspace{0.5cm}{\cal A}_{J+1J+1}=-\frac{1}{2 (J+1)}, \\
&{\cal A}_{J-1J}={\cal A}_{JJ-1}= \frac{\sqrt{(2 J-1)(J+1)}}{2 J},\hspace{0.5cm} {\cal A}_{JJ+1}={\cal A}_{J+1J}=\frac{\sqrt{J (2 J+3)}}{2 J+2} , \\
&{\cal A}_{J-1J+1}={\cal A}_{J+1J-1}=0.
\end{align}

\section{Reduction to $AdS_2$ \label{AdS2}}

The connection of monopole quantization to $AdS_2$ can be made very explicit.
The $\mathbbm{R}^4$ in spherical coordinates is conformally equivalent to $AdS_2\times S^2$:
\begin{equation}
 ds^2=r^2\left(\frac{d\tau ^2+dr^2}{r^2}+d\theta ^2+\sin^2\theta \,d\varphi ^2\right).
\end{equation}
The conformal factor drops out from the action because the scaling symmetry is preserved by the 't~Hooft loop. Expansion in the monopole harmonics can be viewed as the Kaluza-Klein reduction on $S^2$:
\begin{equation}
 \Phi (x)=\frac{1}{r}\sum_{lm}^{}\chi  _{lm}(\tau ,r)Y_{lm}^{(q)}(\theta ,\varphi ),
\end{equation}
resulting in an infinite tower of fields on $AdS_2$.

Take for example the simple scalar with the kinetic operator
\begin{equation}
 -D^2+\frac{B^2}{4r^2}=\frac{1}{r}\left(-\frac{\partial ^2}{\partial \tau ^2}
 -\frac{\partial }{\partial r^2}+\frac{L^2}{r^2}\right)r,
\end{equation}
where $L_i$ is the angular momentum operator in the monopole background. The action for a given KK mode is then
\begin{equation}
 S=\frac{1}{2}\int_{}^{}d\tau dr\,
 \left[
 (\partial _\tau \chi )^2+(\partial _r\chi )^2+\frac{l(l+1)}{r^2}\,\chi ^2
 \right],
\end{equation}
where we took into account the $r^2$ from the measure and used the explicit form of the $L^2$ eigenvalues.
This is just the canonical action of the scalar field of mass $m^2=l(l+1)$ in $AdS_2$. The momentum squared plays the role of the mass-squared operator in the reduced theory.

The scalar field in $AdS_2$ is characterized  by the dimension of the dual operator in the effective $CFT_1$ on the 't~Hooft line, which is related to the mass by the standard formula
\begin{equation}
 \Delta (\Delta -1)=m^2.
\end{equation}
The paramater $\nu $ defined in (\ref{nu-def}) is related to $\Delta $ by
\begin{equation}
 \nu =\Delta -\frac{1}{2}\,.
\end{equation}
For the scalar with $m^2=l(l+1)$, the dimension is $\Delta =l+1$ and $\nu =l+1/2$.

The propagator of the field $\chi $ is just the standard bulk-to-bulk propagator in $AdS_2$: 
\begin{equation}
 G_{\Delta }(\tau _1,r_1;\tau _2,r_2)=\frac{\Gamma (\Delta )\xi ^\Delta }{2^{\Delta +1}\sqrt{\pi }\,\Gamma \left(\Delta +\frac{1}{2}\right)}\,\,
 {}_2\!\,F_1\left(\frac{\Delta }{2}\,,\,\frac{\Delta +1}{2}\,;\Delta +\frac{1}{2}\,;\xi ^2\right),
\end{equation}
where
\begin{equation}
 \xi =\frac{2r_1r_2}{r_1^2+r_2^2+(\tau _1-\tau _2)^2}\,.
\end{equation}

The natural regularization in $AdS_2$ is covariant point-splitting. For $a^2\equiv (\tau _1-\tau _2)^2+(r_1-r_2)^2\rightarrow 0$:
\begin{equation}
 G_\Delta (\xi )\simeq -\frac{1}{2\pi }\left(
 \ln\frac{a}{2r}+\gamma +\Psi (\Delta )
 \right),
\end{equation}
which coincides with $(\ref{regG})$ up to unimportant constant terms that encapsulate regularization ambiguity and should cancel in any properly renormalized (or finite) observable.

\bibliographystyle{nb}

\end{document}